\pdfoutput=1
\PassOptionsToPackage{svgnames,table}{xcolor}
\documentclass[acmlarge,nonacm]{acmart}

\setcopyright{acmlicensed}
\copyrightyear{2026}
\acmYear{2026}
\acmDOI{XXXXXXX.XXXXXXX}

\acmJournal{HEALTH}


\usepackage{algorithmic}
\usepackage[ruled,vlined]{algorithm2e}
\usepackage{url}
\usepackage{multirow}
\usepackage{amsmath}
\usepackage{dirtytalk}
\usepackage{threeparttable}
\usepackage{listings}
\usepackage{pgf-pie}
\usepackage[toc]{appendix}
\usepackage{multicol}
\usepackage{amsfonts}
\usepackage{enumitem}
\usepackage{graphicx}
\usepackage{array}
\usepackage{booktabs}
\usepackage{xspace}
\usepackage{oplotsymbl}
\usepackage{fontawesome5}
\usepackage{lscape}

\usepackage{subfigure}
\usepackage{caption}
\usepackage{wrapfig}
\usepackage{graphicx}
\usepackage[most]{tcolorbox}
\usepackage{adjustbox}
\usepackage{pgfplots}
\pgfplotsset{compat=1.18}
\usepackage{tabulary}
\usepackage{placeins}
\usepackage{needspace}

\usepackage{pgfplotstable}

\pgfplotstableset{
    color cells/.style={
        col sep=comma,
        string type,
        postproc cell content/.code={%
                \pgfkeysalso{@cell content=\rule{0cm}{2.4ex}\cellcolor{black!##1}\pgfmathtruncatemacro\number{##1}\ifnum\number>50\color{white}\fi##1}%
                },
        columns/x/.style={
            column name={},
            postproc cell content/.code={}
        }
    }
}

\definecolor{RED}{RGB}{255,0,0}

\tcbset{
  compactbox/.style={
    colback=gray!10,
    colframe=black,
    boxrule=0.3mm,
    arc=2mm,
    left=1mm,
    right=1mm,
    top=1mm,
    bottom=0.5mm,
    boxsep=0mm,
    before skip=5pt,
    after skip=2pt,
    breakable,
  }
}

\usepackage{tikz}
\usetikzlibrary{calendar,fpu}

\tikzset{
    moon colour/.style={
        moon fill/.style={
            fill=#1
        }
    },
    sky colour/.style={
        sky draw/.style={
            draw=#1
        },
        sky fill/.style={
            fill=#1
        }
    },
    southern hemisphere/.style={
        rotate=180
    }
}

\makeatletter
\pgfcalendardatetojulian{2010-01-15}{\c@pgf@counta}
\def\synodicmonth{29.530588853}
\newcommand{\moon}[2][]{%
    \edef\checkfordate{\noexpand\in@{-}{#2}}%
    \checkfordate%
    \ifin@%
        \pgfcalendardatetojulian{#2}{\c@pgf@countb}%
        \pgfkeys{/pgf/fpu=true,/pgf/fpu/output format=fixed}%
        \pgfmathsetmacro\dayssincenewmoon{\the\c@pgf@countb-\the\c@pgf@counta-(7/24+11/(24*60))}%
        \pgfmathsetmacro\lunarage{mod(\dayssincenewmoon,\synodicmonth)}
        \pgfkeys{/pgf/fpu=false}%
    \else%
        \def\lunarage{#2}%
    \fi%
    \pgfmathsetmacro\leftside{ifthenelse(\lunarage<=\synodicmonth/2,cos(360*(\lunarage/\synodicmonth)),1)}%
    \pgfmathsetmacro\rightside{ifthenelse(\lunarage<=\synodicmonth/2,-1,-cos(360*(\lunarage/\synodicmonth)))}%
    \tikz [moon colour=white,sky colour=black,#1]{
        \draw [moon fill, sky draw] (0,0) circle [radius=1ex];
        \draw [sky draw, sky fill] (0,1ex)
            arc (90:-90:\rightside ex and 1ex)
            arc (-90:90:\leftside ex and 1ex)
            -- cycle;
    }%
}

\def\ACM@receivedfootline{\footnotesize Under review at \@journalNameShort. Received 15 March 2026.}
\AtBeginDocument{%
  \g@addto@macro\ps@firstpagestyle{\fancyfoot[RO,LE]{\ACM@receivedfootline}}%
  \g@addto@macro\ps@standardpagestyle{\fancyfoot[RO,LE]{\ACM@receivedfootline}}%
  \pagestyle{standardpagestyle}%
}

\makeatother

\graphicspath{{./figures/}}

\newcommand\bsub[1]{\addvspace{3pt}\noindent\textbf{#1}}
\newcommand\isub[1]{\noindent\textit{#1}}

\newcolumntype{x}[1]{%
>{\centering\hspace{0pt}}p{#1}}%

\begin{document}

\title[SoK: The Anatomy of Healthcare AI Security and Privacy Research]{SoK: The Anatomy of Healthcare AI Security and Privacy Research — Understanding Threat Taxonomy and Domain Particularities for AI in Healthcare
}

\author{Yuanhaur Chang}
\email{c.yuanhaur@wustl.edu}
\affiliation{%
  \institution{Washington University in St. Louis}
  \city{St. Louis}
  \state{MO}
  \country{USA}
}
\author{Han Liu}
\email{h.liu1@wustl.edu}
\affiliation{%
  \institution{Washington University in St. Louis}
  \city{St. Louis}
  \state{MO}
  \country{USA}
}
\author{Shanghao Shi}
\email{shanghao@wustl.edu}
\affiliation{%
  \institution{Washington University in St. Louis}
  \city{St. Louis}
  \state{MO}
  \country{USA}
}
\author{Jian Xiao}
\email{xjian@wustl.edu}
\affiliation{%
  \institution{Washington University in St. Louis}
  \city{St. Louis}
  \state{MO}
  \country{USA}
}
\author{Chenyang Lu}
\email{lu@wustl.edu}
\affiliation{%
  \institution{Washington University in St. Louis}
  \city{St. Louis}
  \state{MO}
  \country{USA}
}
\author{Ning Zhang}
\email{zhang.ning@wustl.edu}
\affiliation{%
  \institution{Washington University in St. Louis}
  \city{St. Louis}
  \state{MO}
  \country{USA}
}

\begin{abstract}

The integration of artificial intelligence (AI) and machine learning (ML) into healthcare systems holds great promise in improving patient care and care delivery efficiency; however, it also exposes sensitive data and system integrity to potential cyberattacks. Current security and privacy (S\&P) research on healthcare AI is highly unbalanced in terms of healthcare deployment scenarios and threat models, and lacks strong integration with the biomedical research community. This hinders a comprehensive understanding of the risks associated with healthcare AI. To address this gap, we asked 106 healthcare practitioners to evaluate the realism of commonly assumed threat models.
Furthermore, this paper provides a thorough examination of existing healthcare AI S\&P research, offering a unified framework that enables the identification of under-explored areas.
Our survey provides a systematic overview of healthcare AI attacks and defenses, highlighting domain-specificities, practical challenges, and research opportunities for each AI-driven healthcare application domain.
Through our empirical experimental analysis of different threat models and attack/defense feasibility in under-explored healthcare domains, we provide compelling insights into the pressing need for cybersecurity research in the rapidly evolving field of healthcare AI.

\end{abstract}

\maketitle
\renewcommand{\shortauthors}{Chang et al.}

\section{Introduction}
\label{sec:intro}

\bsub{Growing Market of Healthcare AI.}
As technology advances, software systems play an increasingly vital role in commercial products and are becoming integral in the medical field~\cite{ARC_2023}. The development of AI/ML has transformed modern healthcare systems, providing valuable new insights obtained from the vast amount of data collected through patient diagnostics, monitoring, and healthcare research~\cite{Roeloffs_2023}.
It was predicted that the global healthcare AI market would reach 188 billion U.S. dollars by 2030~\cite{Stewart_2023a}. Approximately 22\% of healthcare organizations worldwide stated that they were in the early stage of AI model adoption, while 24\% reported being in the pilot stage~\cite{Stewart_2022}. A recent survey indicated that 44\% of people globally are open to using AI in healthcare for diagnostic and therapeutic purposes~\cite{Stewart_2023}. These highlight the significant benefits of improved diagnostic accuracy and treatment precision for patients, as well as the potential to allow healthcare practitioners to devote more time to patient care instead of routine administrative tasks.

\begin{figure}[t]
  \centering
  \subfigure[Biomedical]{
    \includegraphics[width=0.35\textwidth]{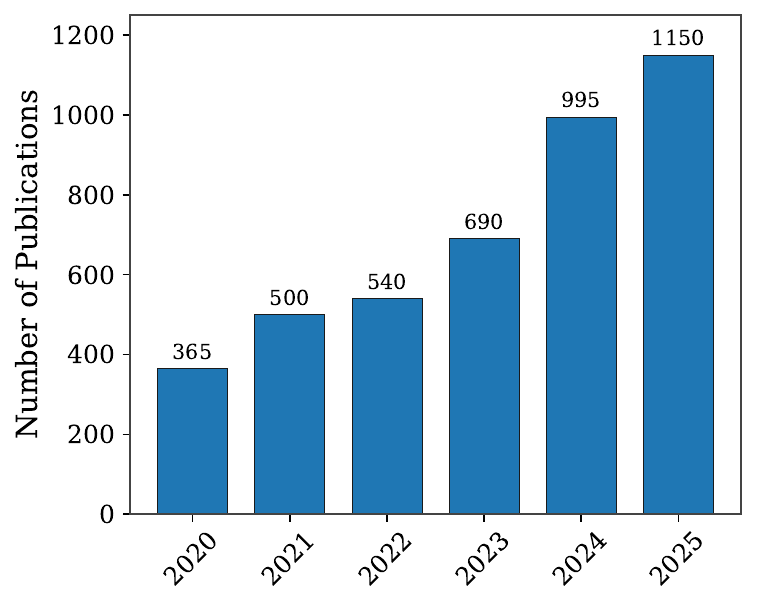}
    \label{fig:bm_pub}
  }
  \subfigure[Security]{
    \includegraphics[width=0.35\textwidth]{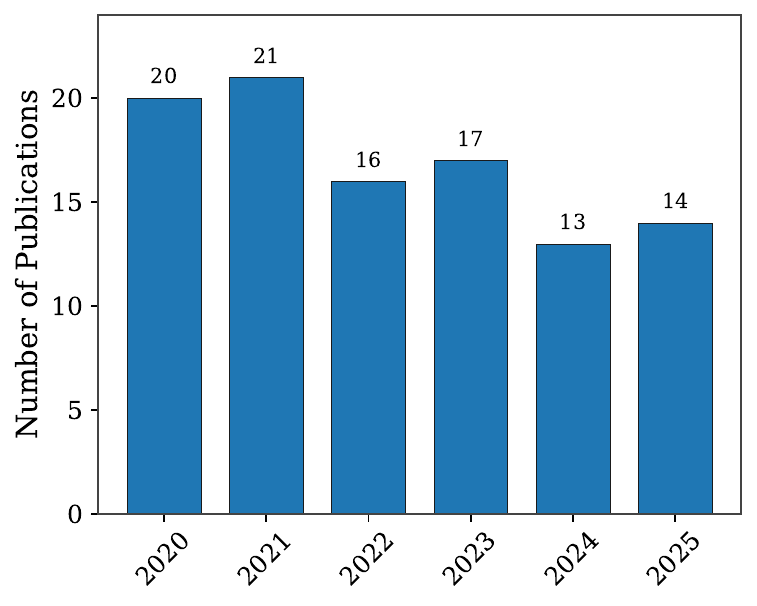}
    \label{fig:sec_pub}
  }
  \caption{Number of healthcare AI publications from the biomedical and security communities from 2020 to 2025.}
  \label{fig:pubstats}
\end{figure}

\bsub{Gaps in Healthcare AI S\&P Research.}
While healthcare AI research has received unprecedented attention among biomedical communities, there is a limited amount of understanding of the adversarial robustness of this technology in the context of healthcare applications. Among all the papers we surveyed from both computer science venues (including security and AI/ML venues) and biomedical venues, there is a limited amount of research work focusing on these systems, with roughly 10--20 publications each year (Figure~\ref{fig:pubstats}).
Besides the difference in research volume, there is also a disconnect between the security and the biomedical communities in the assumed threat models and technologies under study. Given the increasing role AI plays in delivering healthcare solutions, we aim to identify this gap and contextualize the risks of AI applications in different healthcare domains.

\bsub{Our Methodology.}
To understand where healthcare AI is being studied and deployed, we collected more than 4,200 articles in the past 6 years (2020--2025) from major medical journals (including but not limited to Nature, Science, Cell, The Lancet, and The New England Journal of Medicine).
Drawing from this exercise, we developed the systematization of healthcare applications of AI, based on which it then becomes possible to contextualize the threat model for the analysis of the security works.
To understand the current state of the art in security and privacy for healthcare AI, we collected a corpus of 109 articles from both selected security venues and AI venues. Building on these articles, we not only systematize common threat models but also investigate whether healthcare practitioners' perspectives and experiences align with them.
We create a taxonomy of existing attacks and defenses based on healthcare application domains (Table~\ref{tab:systemization_all_attacks} and Table~\ref{tab:systemization_all_defense}). Through this exercise, we highlight domain-specific challenges and areas where there is limited understanding, and discuss future directions to safeguard these healthcare technologies.
To pioneer the research opportunities in under-explored domains, we conducted experimental validations considering realistic threat models in healthcare.

\begin{figure*}[t]
\centering
\includegraphics[width=1.0\textwidth]{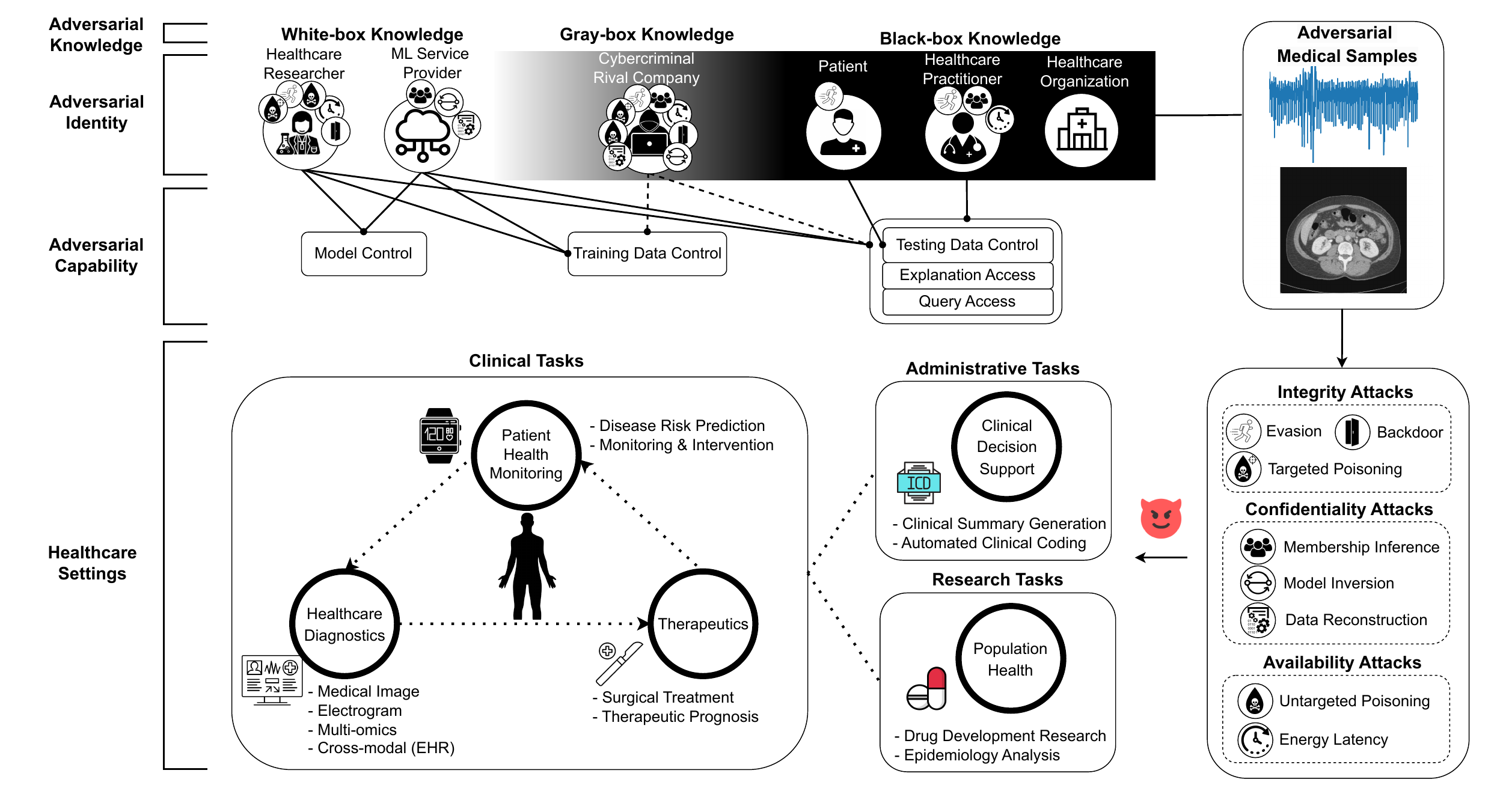}
\caption{Adversarial knowledge, capability, and goals based on the adversary's identity in the healthcare setting.}
\label{fig:overview}
\end{figure*}

\bsub{Contributions.}
While there have been works systematizing AI/ML research in network intrusion detection~\cite{nadeem2023sok}, security applications~\cite{apruzzese2023sok}, explainability~\cite{noppel2023sok}, accountability~\cite{birhane2024sok}, and privacy~\cite{salem2023sok}, a unified view on the topic in healthcare settings is not yet available.
To bridge the gaps, we aim to present a systematic survey unifying existing works and propose a comprehensive framework that helps shed light on challenges and opportunities for future research. Our SoK makes the following contributions:
\begin{itemize}[topsep=1pt]
    \setlength\itemsep{0em}
    \item We contextualize traditional AI attack threat models with the healthcare setting, providing a taxonomy of threats in healthcare, including common adversarial identities, knowledge, capabilities, and goals (Section~\ref{sec:threat_models}, Figure~\ref{fig:overview}). We further conduct a case study with 106 healthcare practitioners to investigate whether these threat models align with their experiences in the field.

    \item We provide a comprehensive systematization of current S\&P research on healthcare AI in various application domains, pointing out challenges and opportunities for future research (Sections~\ref{sec:Medical_Diagnostics_Systems}--\ref{sec:Patient_Health_Monitoring}, Table~\ref{tab:systemization_all_attacks}, Table~\ref{tab:systemization_all_defense}).

    \item We conducted proof-of-concept validation with tailored experiments in diverse, under-explored healthcare domains. We specifically sought to explore the S\&P risks of generative AI models in healthcare, along with traditional ML methods and deep neural networks (DNN), through experiments to validate the adversarial robustness of emerging healthcare AI (Section~\ref{sec:exp}).
\end{itemize}

\section{Literature Survey Methods}
\label{sec:method}

To scope to studies on security and privacy risks of healthcare AI, we excluded research that only explores the applicability of AI in healthcare tasks. We also excluded those whose research goal was to improve healthcare AI's accuracy, efficiency, or other performance metrics. We included works in our collected corpus if (i) the paper studied the security or privacy issue of healthcare AI, and (ii) it conducted attack or defense experiments on at least one healthcare dataset.

\bsub{Identifying Relevant Work.}
For our focus on the security and privacy issues of healthcare AI, we defined unique search strings with a combination of keywords related to the topic.
We searched for relevant papers in the past 10 years in notable security (S\&P, Security, CCS, NDSS, AsiaCCS, EuroS\&P, SOUPS, PoPETS), AI (AAAI, CVPR, NeurIPS, ICML, KDD), and medical (Nature, Nature Communications, Science, Cell, Lancet, NEJM, MICCAI, JMIR) venues. As relevant studies are also published beyond these venues, we also searched Google Scholar to find relevant papers in broader computing journals (e.g., Sensors, Electronics).
Three researchers independently read the abstract, introduction, and methodology sections of each candidate paper to determine whether the paper meets our inclusion criteria.

\bsub{Qualitative Analysis.}
We qualitatively analyzed papers in our dataset by coding the healthcare application domains the study falls in and the security and privacy issues they explored. The systematization was created with iterative coding sessions between two researchers. First, two researchers inductively coded all papers independently. The researchers then met to resolve inconsistencies and make adjustments iteratively. After adjustments were made, one researcher recoded the remaining dataset. The process continued until both researchers reached an agreement on the final systematization.

\bsub{Dataset Statistics.}
The final corpus contains 109 healthcare AI S\&P works.
Our literature survey reveals several interesting insights into the current state of S\&P research in healthcare AI. First, the current research is skewed toward certain threat types. For example, about 39\% of surveyed studies focused on adversarial examples for evasion attacks, while only about 7\% addressed issues related to AI model availability. Furthermore, the healthcare domain being studied is highly unbalanced. About 36\% of healthcare AI S\&P research focused on image-related diagnostic tasks, while other application domains, such as disease risk prediction, are largely overlooked (2\%). Another emerging trend is the rapid rise of generative AI in healthcare research, which has gained increasing attention in the biomedical community since 2020. Although generative models raise new security risks for healthcare providers, the security community has largely overlooked these concerns (Appendix~\ref{app:stats}). These gaps point to numerous opportunities for future research, underscoring the need to extend security and privacy investigations into a broader range of healthcare AI application domains.

\section{Taxonomy of Threats in Healthcare}
\label{sec:threat_models}

\subsection{Adversarial Identity}
\label{subsec:adversary_identity}

We identify the adversary's potential identity in the healthcare domain and their respective motivations. However, it should be noted that in the real world, multiple adversaries can work together to achieve malicious intent. Their adversarial capability and knowledge may, therefore, be expanded and need to be discussed case-by-case.

\bsub{Patient.}
Patients generally would not have direct access to healthcare AI. However, except for data that come directly from hospital measurements, there are cases where they are provided by the patients themselves, especially with the rise of telemedicine. Such patient-provided healthcare data is vulnerable to \textit{manipulation}~\cite{Hayward_2024}. Malicious patients can be financially motivated to manipulate data for insurance fraud.
They may manipulate self-reported or patient-generated inputs to bias diagnostic outputs toward false-negative results to avoid social stigma associated with certain medical conditions, or for secondary gain, such as obtaining sick leave or time off work.
In addition, with the increasing use of LLM-based medical AI assistants in telemedicine, patients may also act as \textit{prompt-level adversaries} by using jailbreak prompts and fraudulent supporting content to bypass safety safeguards. For example, recent news reporting suggests that a medical AI model could be manipulated through relatively simple jailbreak techniques to produce harmful or misleading prescription-related recommendations~\cite{sabin2026doctronic}.
Even non-malicious users may inadvertently trigger unsafe responses through ambiguous prompts, which is especially concerning in therapeutic or mental-health settings.

\bsub{Healthcare Practitioner.}
Healthcare practitioners include clinicians, nurses, or any other person working at or near the point of care. Unlike patients, healthcare practitioners operate as \textit{insiders} in the clinical workflow, directly interacting with healthcare AI systems and typically having access to both the initial input data and the preliminary model outputs.
Malicious healthcare practitioners can be motivated to manipulate model outputs to justify overutilization, leading to unnecessary tests, procedures, or treatments and additional healthcare spending.
They may also conduct healthcare insurance fraud, such as phantom billing, which bills services or supplies that the patients never actually received~\cite{kose2015interactive}. For example, the U.S. Department of Justice recently charged 193 defendants, including 76 doctors, nurse practitioners, and other medical professionals across 32 federal districts in a healthcare fraud scheme that resulted in \$1.6 billion in actual losses~\cite{Justice_2024}.

\bsub{Cloud-based ML Service Provider.}
Hospitals may opt for cloud-based ML service instead of on-site deployment to minimize investments in hardware and related IT infrastructures~\cite{GHX_2023}. Adversarial service providers are typically considered \textit{honest-but-curious}~\cite{backes2017identifying} by the community, and may have capabilities that range from training data control to model control. They usually aim to extract private patient records and put data confidentiality at risk.

\bsub{Third-Party Healthcare Organization.}
The profit and service of organizations such as healthcare insurance companies may depend directly on the outcome of healthcare AI predictions. Thus, they may have the highest monetary incentives to manipulate healthcare AI~\cite{kose2015interactive,Tong_2024}. They typically have limited capability, knowledge, and access to the models, and are more likely to \textit{work with other adversaries}.

\bsub{Healthcare Researcher.}
Researchers who collect original datasets and design foundation models for downstream use may be in the best position to compromise model integrity and availability~\cite{saini2022predatory}. Because they operate upstream in the medical AI pipeline, they may introduce biased or manipulated data, or even release backdoored models that \textit{affect many downstream users}. For instance, pharmaceutical companies may have an interest in corrupting clinical trials and studies in the context of groundbreaking drug research~\cite{Elliott_2010}.

\bsub{Rival Service Provider.}
Rival companies can leverage domain expertise to perform gray-box attacks, \textit{targeting system availability} to diminish model performance or shut down AI diagnostic systems to discredit competing businesses. Moreover, as models themselves become valuable assets, rival companies may leverage \textit{model-stealing attacks} to extract competitors' proprietary models and compromise intellectual property.

\bsub{Cybercriminal.}
Although cybercriminals typically lack direct model or training data control, they often \textit{exploit known vulnerabilities} in common healthcare systems, such as Picture Archiving and Communication Systems (PACS) and legacy OSes (e.g., Windows XP or unpatched Windows Server) that are still in use by many hospitals~\cite{Eastwood_2024} to deploy ransomware~\cite{mirsky2019ct}. In addition, cybercriminals could use social engineering and spear-phishing to \textit{install malware}, which is what prevented Illinois hospitals from submitting insurance claims to Medicaid, complicating their finances~\cite{Collier_2023}.
Moreover, as emerging agentic AI systems become more widely used, cybercriminals may inject malicious instructions into third-party resources (e.g., webpages, documents, or databases) that are later retrieved by the agent, thereby performing \textit{indirect prompt injection}. This emerging new attack can manipulate medical agent behavior, leading to harmful clinical or operational consequences~\cite{greshake2023not}.
Besides financial incentives, adversarial nation-states may choose to cause harm to government officials. For instance, former US Vice President Dick Cheney reportedly had the wireless functionality of his pacemaker disabled to mitigate the risk of terrorist interference~\cite{Franzen_2013}.

\subsection{Adversarial Capability}
\label{subsec:adversary_capability}

\bsub{Training Data Control.} Adversaries may gain partial control over the training data, enabling them to insert or modify training samples. This is a foundational strategy in data poisoning attacks. Particular to clean-label attacks is the constraint on label manipulation, where adversaries cannot assign labels to the poisoned samples, unlike regular poisoning attacks that assume label control.

\bsub{Model Control.} Adversaries may take control of the model parameters through various means, such as embedding a Trojan trigger within the model parameters or executing malicious updates in federated learning scenarios.

\bsub{Testing Data Control.} At the model deployment stage, attackers can introduce perturbations to inference samples. This is required for evasion and adversarial examples.

\bsub{Query Access.}
Adversaries can query the model for predictions, which can include both labels and confidence scores. This capability is required for black-box evasion attacks, energy-latency attacks, and privacy attacks.

\bsub{Explanation Access.}
This capability is especially related to healthcare systems that employ explainable ML. Adversaries who obtain access to the explanations generated by these tools can exploit this information not only to undermine the integrity of the explanations themselves but also to enhance the effectiveness of other attacks.

\bsub{Third-Party Data Injection.} Adversaries may passively inject malicious content into third-party resources (e.g., webpages, documents, or databases) that are later retrieved and incorporated by the target system. This capability is essential for indirect prompt injection attacks against emerging medical agentic AI systems.

\subsection{Adversarial Knowledge}

\bsub{White-box Access.} The adversary has full knowledge of the target system, including the model architecture, parameters, training objective, and defense mechanisms. In some settings, this may also include knowledge of preprocessing pipelines, prompt templates, or agent policies/tool descriptions.

\bsub{Gray-box Access.} The adversary has partial knowledge of the target system but lacks complete access to internal parameters or implementation details. For example, the adversary may know the model family, task setting, data modality, or preprocessing pipeline, and may use auxiliary datasets or surrogate models to approximate the target system.

\bsub{Black-box Access.} The adversary has no knowledge of the model internals and can only infer information from observable behavior, such as prediction outputs (e.g., labels or confidence scores) and system responses. In this setting, attacks rely on querying or external observations, rather than direct knowledge of the model design.

\subsection{Adversarial Goals}

\bsub{Integrity Attacks.}
Integrity attacks aim to cause healthcare AI systems to deviate from their intended behavior, resulting in adversarial outcomes such as incorrect predictions, unsafe responses, or improper system actions. Depending on the adversary's capabilities and the stage of attack, these attacks include inference-time evasion attacks (e.g., adversarial examples)~\cite{bortsova2021adversarial}, training-time poisoning attacks (e.g., targeted, backdoor, and clean-label poisoning)~\cite{sarkar2021trapdoor}, parameter-level model poisoning, and emerging LLM-specific attacks such as jailbreaks and prompt injection against medical AI assistants and agentic systems. The threat model for integrity attacks depends on the attack stage. Evasion attacks are typically mounted in black-box or gray-box settings (e.g., score-/decision-based attacks, or attacks using surrogate models and auxiliary datasets)~\cite{bortsova2021adversarial}, whereas poisoning attacks assume partial control over the training data and, in some cases, limited knowledge of the training pipeline~\cite{sarkar2021trapdoor}. Parameter-level model poisoning requires direct control over model updates or parameters, while LLM-specific integrity attacks typically assume query access and/or the ability to influence retrieved third-party content.

\bsub{Confidentiality Attacks.}
Confidentiality attacks aim to extract sensitive information from healthcare AI systems, including proprietary models and confidential training data. These attacks can be categorized into \emph{model privacy attacks} (e.g., model extraction or model stealing)~\cite{fredrikson2014privacy} and \emph{data privacy attacks} (e.g., membership inference and reconstruction attacks)~\cite{berrang2018dissecting}. A common threat model assumes that the adversary knows the target model architecture (and possibly the training procedure) and has access to an auxiliary dataset drawn from a similar distribution, which can enable effective privacy attacks in both gray-box and white-box settings.

\bsub{Availability Attacks.}
Availability attacks aim to disrupt the normal operation of healthcare AI systems by degrading service reliability or rendering services unusable, for example, by reducing overall model performance, increasing inference latency, or causing excessive resource consumption. These attacks can be categorized into \emph{performance-degradation attacks} (e.g., indiscriminate poisoning) and \emph{resource-exhaustion attacks} (e.g., energy-latency attacks)~\cite{liu2023slowlidar}. A common threat model assumes that the adversary can influence a portion of the training data (for poisoning-based attacks) or query the deployed model (for latency/energy attacks), often under gray-box or black-box settings; in gray-box settings, adversaries may further leverage auxiliary models and transferability to generate effective poisoning samples that degrade overall model performance~\cite{demontis2019adversarial,suciu2018does}.

\subsection{Case Study: Healthcare Practitioners' Perspective}

\bsub{Overview.}
To assess how healthcare practitioners' perceptions and experiences align with threat models commonly assumed in healthcare AI security research, we conducted a structured survey of professionals across diverse roles and healthcare institutions. We aim to answer the following question:
\textit{To what extent do healthcare practitioners' perceptions and experiences with real-world security incidents align with the threat models assumed in healthcare AI S\&P research?}

\begin{table}[t]
\centering
\scriptsize
\caption{Primary Role of Surveyed Healthcare Practitioners.}
\label{tab:demographic}
\resizebox{0.7\linewidth}{!}{%
\begin{tabular}{llll}
\specialrule{.2em}{.1em}{.1em}
\multicolumn{2}{l}{\textbf{Primary Role in Healthcare}} & \multicolumn{2}{l}{\textbf{Years of Experience}} \\
\specialrule{.2em}{.1em}{.1em}
Physician/Doctor & 12.3\% & Less than 1 year & 1.9\% \\
Nurse & 24.5\% & 1--5 years & 36.8\% \\
Healthcare IT Administrator & 0.9\% & 6--10 years & 21.7\% \\
Clinical Researcher & 9.4\% & More than 10 years & 39.6\% \\
Pharmacist & 15.1\% &  &  \\
Therapist & 13.2\% &  &  \\
Healthcare Administrator/Manager & 15.1\% &  &  \\
Medical Assistant & 7.5\% &  &  \\
Other & 1.9\% &  &  \\ \specialrule{.2em}{.1em}{.1em}
\end{tabular}%
}
\end{table}

\bsub{Participant Recruitment.}
We recruited 106 healthcare professionals in the US from Prolific\footnote{\url{https://www.prolific.com}}, with a minimum prior survey approval rate of 90\%. The study took approximately 15 minutes to complete.
Our study was reviewed by the university IRB and deemed exempt. In Table~\ref{tab:demographic}, we summarize the demographic statistics of our survey participants.

\bsub{Study Design.}
We evaluate practitioners' assessments of healthcare AI security threats along three dimensions: (i) perceived realism of specific threat scenarios, (ii) perceived feasibility of adversaries obtaining certain adversarial capabilities, and (iii) level of concern regarding attack goals. Participants are also invited to report any personal experiences with AI-related security incidents and describe incidents that fall outside the predefined categories.

\subsection{Study Findings}

\begin{figure}[tbp]
\centering
\includegraphics[width=0.6\columnwidth]{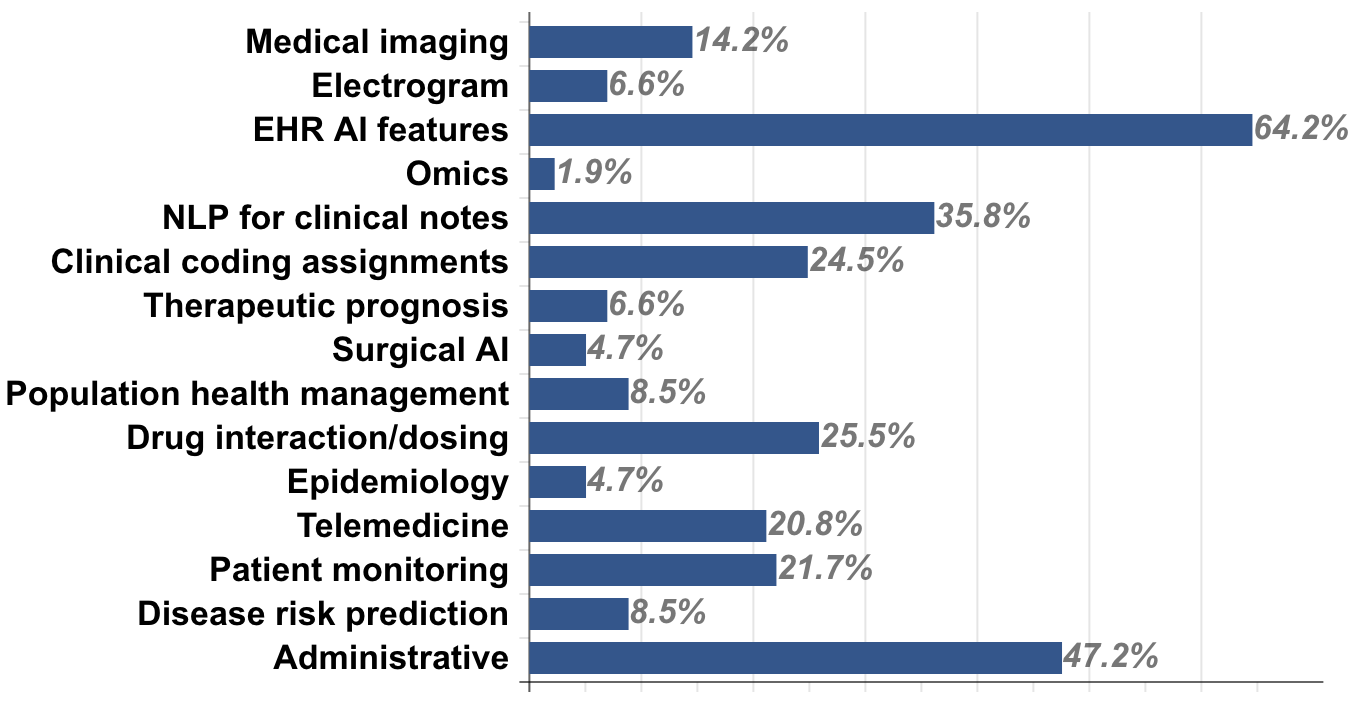}
\caption{Participants' view of how healthcare AI is used.}
\label{fig:usage}
\end{figure}

\bsub{Healthcare AI is used mainly in routine clinical and administrative workflows.} Figure~\ref{fig:usage} shows healthcare practitioners' views on where healthcare AI is used in practice, since the deployment context helps determine which threats are most realistic. The results suggest that current use is concentrated in electronic health record (EHR)-related features (64.2\%) and administrative workflows (47.2\%), with additional adoption in NLP for clinical notes, drug interaction and dosing support, and clinical coding. By contrast, relatively few participants reported AI use in more specialized domains such as medical imaging, prognosis, surgical AI, epidemiology, or omics. A key takeaway from practitioners' responses is that healthcare AI is used primarily in routine tasks, rather than independent clinical decision-making systems. From their perspective, many realistic risks therefore arise in everyday operational systems. In these settings, attacks on data integrity, privacy, or service availability can directly interfere with normal healthcare workflows.

\begin{figure}[tbp]
\centering
\includegraphics[width=0.75\columnwidth]{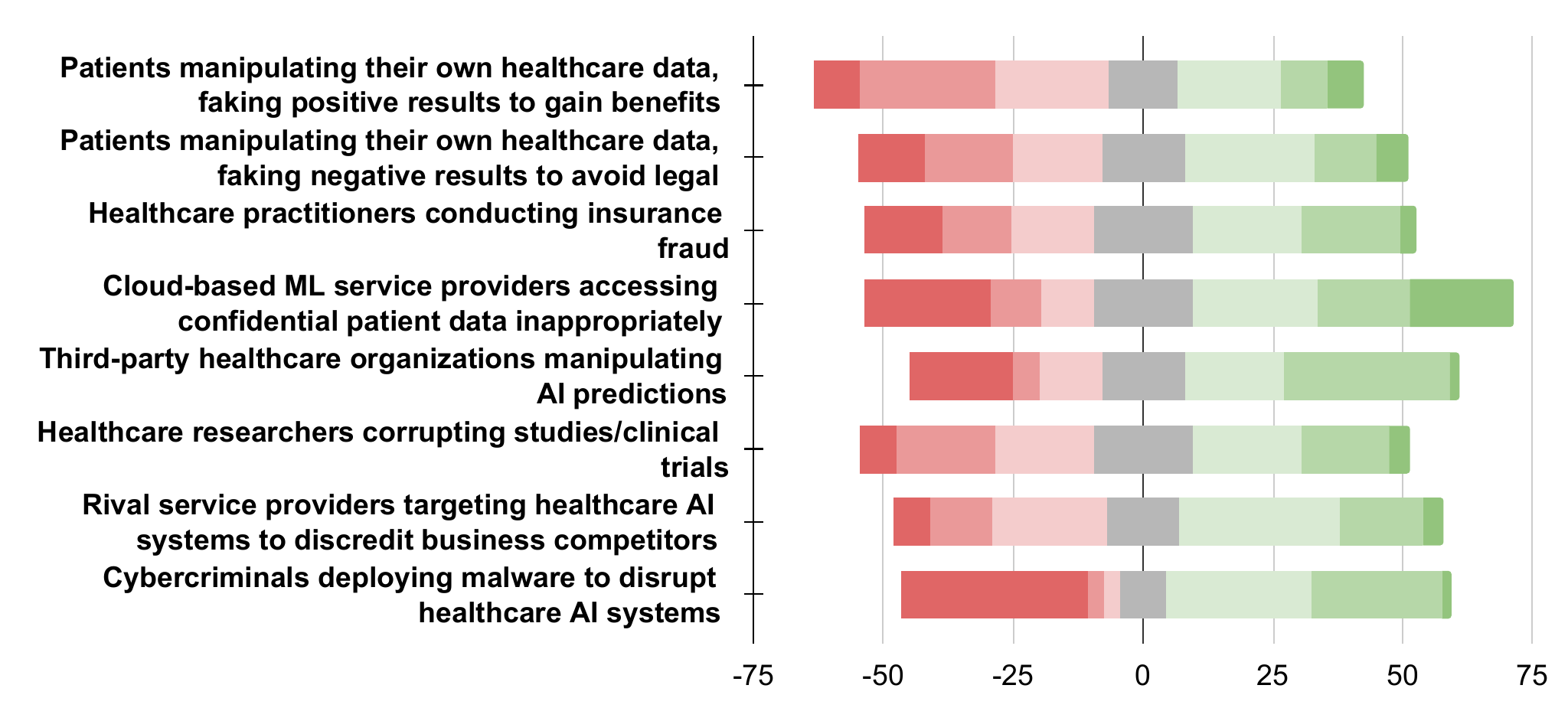}
\caption{Participants' realism rating on predefined healthcare AI threat scenarios.}
\label{fig:realism}
\end{figure}

\bsub{Threat scenarios involving external actors and internal organizational actors are widely viewed as realistic.} Figure~\ref{fig:realism} examines how practitioners assess the realism of threat scenarios associated with the adversarial identities discussed in \S\ref{subsec:adversary_identity}, since these judgments help indicate which threat models are most relevant in practice. The results show that several scenarios were seen as plausible in real healthcare settings, especially those involving cloud-based ML service providers improperly accessing patient data, third-party healthcare organizations manipulating AI predictions, rival service providers targeting healthcare AI systems, and cybercriminals disrupting healthcare AI systems through malware. The key finding is that practitioners see both external attackers and organizations already involved in healthcare AI deployment as realistic threats. This means the threat model is not limited to external attackers. In practice, risks may also come from service providers or partner organizations that already have access to systems, data, or operational processes.

\bsub{Perceived adversarial capability differs sharply by adversarial identity.} Table~\ref{tab:feasibility} shows how practitioners assess the feasibility of the adversarial capabilities described in \S\ref{subsec:adversary_capability}, since these judgments help indicate which adversaries deserve the most attention in the threat model. The results show a clear divide across adversarial identities. Patients were consistently viewed as the least capable adversary, with low ratings for accessing or modifying training data, directly accessing model parameters or architecture, and altering healthcare data used as model input. By contrast, cybercriminals and cloud ML service providers were seen as much more capable, especially in training data access or manipulation, model access, and input data manipulation. Healthcare practitioners, healthcare organizations, and healthcare researchers were also rated relatively higher for obtaining AI recommendations and accessing explanations, likely because they work more closely with these systems. The key finding is that practitioners associate adversarial capability with technical access and organizational position, rather than with low-access end users. For the threat model, this means that more attention should be given to actors who already have access to systems, data, or infrastructure, because they are seen as much more capable of launching attacks.

\begin{table*}[tbp]
\centering
\scriptsize
\setlength{\tabcolsep}{5pt}
\caption{Participants' feasibility ratings on adversarial identity vs. capability.}
\label{tab:feasibility}
\begin{tabulary}{\linewidth}{lCCCCCCC}
\specialrule{.2em}{.1em}{.1em}
 & \multicolumn{7}{c@{}}{Adversarial Identity} \\ \cline{2-8}
& Patient & \begin{tabular}[c]{@{}c@{}}Healthcare \\ Practitioner\end{tabular} & \begin{tabular}[c]{@{}c@{}}Cloud ML \\ Provider\end{tabular} & \begin{tabular}[c]{@{}c@{}} Healthcare \\ Organization\end{tabular} & \begin{tabular}[c]{@{}c@{}}Healthcare \\ Researcher\end{tabular} & \begin{tabular}[c]{@{}c@{}}Rival Service \\ Provider\end{tabular} & Cybercriminal \\ \hline
\begin{tabular}[c]{@{}l@{}}Able to access, insert, or alter samples within \\ healthcare datasets used for AI model training\end{tabular} & \cellcolor[HTML]{91CFC9}2.19 & \cellcolor[HTML]{FDB8A6}4.78 & \cellcolor[HTML]{FDAD98}4.92 & \cellcolor[HTML]{FFEAE4}4.16 & \cellcolor[HTML]{FFEAE4}4.10 & \cellcolor[HTML]{EEF8F7}3.56 & \cellcolor[HTML]{FD9184}5.14 \\ \hline
\begin{tabular}[c]{@{}l@{}}Direct access to AI model parameters \\or architecture\end{tabular} & \cellcolor[HTML]{81C7C1}1.96 & \cellcolor[HTML]{D9EFEC}3.25 & \cellcolor[HTML]{FEC0AE}4.69 & \cellcolor[HTML]{F0F8F8}3.59 & \cellcolor[HTML]{FCFDFD}3.75 & \cellcolor[HTML]{EEF8F7}3.52 & \cellcolor[HTML]{FDAD98}4.94 \\ \hline
\begin{tabular}[c]{@{}l@{}}Able to alter healthcare data that is input \\ to the healthcare AI systems for prediction\end{tabular} & \cellcolor[HTML]{91CFC9}2.40 & \cellcolor[HTML]{FECABD}4.58 & \cellcolor[HTML]{FECABD}4.53 & \cellcolor[HTML]{FFF0EC}4.01 & \cellcolor[HTML]{FED8CE}4.34 & \cellcolor[HTML]{EEF8F7}3.56 & \cellcolor[HTML]{FD9184}5.18 \\ \hline
\begin{tabular}[c]{@{}l@{}}Ability to obtain healthcare recommendations \\ or decisions from AI systems\end{tabular} & \cellcolor[HTML]{FDAD98}4.93 & \cellcolor[HTML]{FC7957}5.65 & \cellcolor[HTML]{FDBAA9}4.75 & \cellcolor[HTML]{FCA289}5.02 & \cellcolor[HTML]{FD9184}5.13 & \cellcolor[HTML]{FED1C5}4.44 & \cellcolor[HTML]{FD9174}5.32 \\ \hline
\begin{tabular}[c]{@{}l@{}}Access to explanations of why the AI made \\ specific healthcare recommendations or decisions\end{tabular} & \cellcolor[HTML]{FFF8F7}3.90 & \cellcolor[HTML]{FCA289}5.09 & \cellcolor[HTML]{FDB19D}4.82 & \cellcolor[HTML]{FEC0AE}4.69 & \cellcolor[HTML]{FD9184}5.18 & \cellcolor[HTML]{FFEAE5}4.09 & \cellcolor[HTML]{FDB19D}4.88 \\ \specialrule{.2em}{.1em}{.1em}
\end{tabulary}%
\end{table*}

\begin{figure}[tbp]
\centering
\includegraphics[width=0.6\columnwidth]{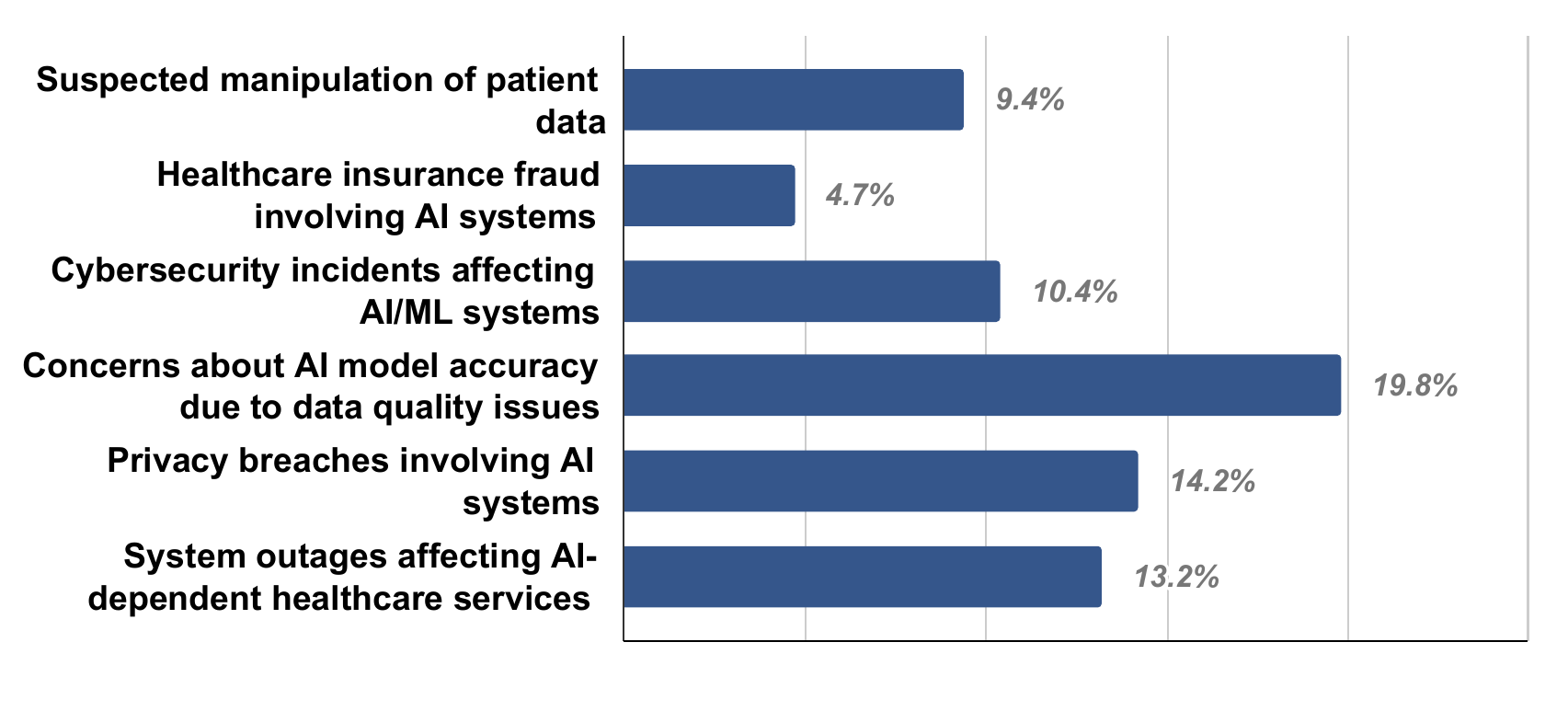}
\caption{Participants' concerns regarding healthcare AI attack goals.}
\label{fig:incidents}
\end{figure}

\bsub{Practitioners' concerns align with integrity, confidentiality, and availability, but are expressed in operational terms.} Figure~\ref{fig:incidents} shows practitioners' concerns about different types of healthcare AI incidents, since these concerns help show how core threat model goals appear in real deployment settings. The results show that the most frequently selected concern was AI model accuracy issues due to data quality problems, which points to concerns about the integrity of healthcare AI outputs. Privacy breaches in AI systems relate to confidentiality, while service outages and cyberattacks on AI/ML systems raise concerns about availability and operational resilience. The key finding is that practitioners think about healthcare AI security mainly in terms of its practical impact on care delivery, patient privacy, and service continuity, rather than in terms of specific model-level attack techniques. This suggests that the threat model should connect traditional security goals such as integrity, confidentiality, and availability to the practical consequences that matter in real healthcare settings.

\FloatBarrier

\section{Healthcare Diagnostics Systems} \label{sec:Medical_Diagnostics_Systems}

Healthcare AI in patient diagnostics is rapidly revolutionizing the medical field, with AI-driven tools capable of detecting subtle patterns in medical images, genetic data, and patient records at or above the level of experienced clinicians.
However, with the involvement of various forms of patient data also come security and privacy concerns. In this section, we discuss diagnostics AI that takes different data modalities as inputs, including images (\S\ref{subsec:image}), electrograms (\S\ref{subsec:electrogram}), multi-omics data (\S\ref{subsec:multiomics}), and cross-modal data (\S\ref{subsec:crossmodal}).

\begin{table}
\rotatebox{90}{%
\begin{minipage}{\textheight}
\caption{Systematization of existing integrity, confidentiality, and availability attacks in the healthcare setting.}
\label{tab:systemization_all_attacks}
\footnotesize
\centering
\setlength{\tabcolsep}{3.5pt}
\begin{adjustbox}{max width=\linewidth}
\begin{tabular}{cccccccccc}
\specialrule{.2em}{.1em}{.1em}
\multicolumn{3}{c}{\multirow{2}{*}{}} & \multicolumn{3}{c}{Integrity} & \multicolumn{3}{c}{Confidentiality} & \multirow{2}{*}{Availability} \\ \cline{4-9}
\multicolumn{3}{c}{} & Evasion & Poison & Backdoor & \begin{tabular}[c]{@{}c@{}}Membership \\ Inference\end{tabular} & \begin{tabular}[c]{@{}c@{}}Model \\ Inversion\end{tabular} & \begin{tabular}[c]{@{}c@{}}Data \\ Reconstruction\end{tabular} &  \\
\specialrule{.2em}{.1em}{.1em}
\multirow{6}{*}{\begin{tabular}[c]{@{}c@{}}Healthcare \\ Diagnostics\end{tabular}} & \multirow{3}{*}{\begin{tabular}[c]{@{}c@{}}Image \\ Diagnostic\end{tabular}} & Classification & \begin{tabular}[c]{@{}c@{}}\squad~\faFile*~\cite{ma2021understanding,lal2021adversarial}\\ \squadfill~\faFile*~\cite{zhou2021machine,bortsova2021adversarial}\\ \squadfill~\faFile*~\faQuestionCircle ~\cite{paschali2018generalizability}\\ \squad~\squadfill~\faFile*~\faQuestionCircle ~\cite{finlayson2018adversarial,dong2024survey}\\ \squad~\faDatabase~\faFile*~\faRobot~\cite{rosenblatt2023gradient}\end{tabular} & \begin{tabular}[c]{@{}c@{}}\faDatabase ~\cite{martinelli2023data}\\ \squadfill~\faDatabase ~\cite{mozaffari2014systematic,singkorapoom2023pre,sun2021data}\\ \squadfillhl~\faDatabase~\faRobot ~\cite{kalapaaking2023blockchain}\end{tabular} & \begin{tabular}[c]{@{}c@{}}\squadfill~\faDatabase ~\cite{nwadike2020explainability,feng2022fiba}\\ \squad~\faDatabase~\faRobot ~\cite{jinbackdoor}\end{tabular} & \squad~\squadfill~\squadfillhl~\faQuestionCircle ~\cite{gupta2021membership} & \squadfill~\faQuestionCircle ~\cite{wu2020evaluation} & \moon{8} & \begin{tabular}[c]{@{}c@{}}\squadfill~\faDatabase ~\cite{sun2024medical}\\ \squad~\faDatabase~\faRobot ~\cite{jinbackdoor}\end{tabular} \\ \cline{3-10}
 &  & Detection & \begin{tabular}[c]{@{}c@{}}\squadfill~\faFile*~\faQuestionCircle ~\cite{mangaokar2020jekyll} \\ \squadfill~\faFile*~\cite{yao2020miss}\end{tabular} & \moon{15} & \begin{tabular}[c]{@{}c@{}}\squadfill~\faDatabase ~\cite{feng2022fiba}\\ \squad~\faDatabase~\faRobot ~\cite{matsuo2021backdoor}\end{tabular} & \moon{15} & \moon{15} & \moon{8} & \moon{15} \\ \cline{3-10}
 &  & Segmentation & \begin{tabular}[c]{@{}c@{}}\squadfill~\faFile*~\faQuestionCircle ~\cite{ozbulak2019impact, shao2021target,paschali2018generalizability}\\ \squad~\squadfill~\faFile*~\faQuestionCircle ~\cite{wang2022feature,dong2024survey}\end{tabular} & \moon{15} & \squadfill~\faDatabase ~\cite{feng2022fiba} & \squadfillhl~\faDatabase~\faQuestionCircle ~\cite{chobola2023membership} & \squadfillhl~\faQuestionCircle ~\cite{subbanna2021analysis} & \moon{8} & \squadfill~\faDatabase ~\cite{lin2024safeguarding} \\ \cline{2-10}
 & \multicolumn{2}{c}{Electrogram Diagnostics} & \begin{tabular}[c]{@{}c@{}}\squad~\faFile*~\cite{han2020deep,ono2021application}\\ \squadfill~\faFile*~\cite{aminifar2020universal}\\ \squadfill~\faFile*~\faQuestionCircle ~\cite{chen2020ecgadv,lam2020hard}\end{tabular} & \squad~\faDatabase ~\cite{ismail2023analyzing} & \moon{8} & \moon{15} & \moon{15} & \squadfillhl~\faQuestionCircle ~\cite{abuadbba2020can} & \squad~\faDatabase ~\cite{ismail2023analyzing} \\ \cline{2-10}
  & \multicolumn{2}{c}{Multi-omics Diagnostic} & \begin{tabular}[c]{@{}c@{}}\squadfill~\faFile*~\faQuestionCircle ~\cite{skovorodnikov2024fimba,montserrat2023adversarial}\\ \squad~\squadfill~\faFile*~\faQuestionCircle ~\cite{ghaffari2022adversarial}\end{tabular} & \moon{15} & \squad~\faDatabase ~\cite{sarkar2021trapdoor} & \begin{tabular}[c]{@{}c@{}}\squad~\faQuestionCircle ~\cite{chen2020differential}\\ \squadfillhl~\faQuestionCircle ~\cite{hagestedt2020membership}\end{tabular} & \moon{15} & \squadfillhl~\faQuestionCircle~\cite{pan2020privacy,berrang2018dissecting} & \moon{15} \\ \cline{2-10}
 & \multicolumn{2}{c}{Cross-modal Diagnostic} & \begin{tabular}[c]{@{}c@{}}\faFile*~\cite{sun2018identify}\\ \squad~\squadfillhl~\faFile*~\faQuestionCircle ~\cite{an2019longitudinal}\end{tabular} & \moon{15} & \faDatabase ~\cite{joe2021machine} & \squadfill~\faQuestionCircle ~\cite{zhang2022membership} & \moon{15} & \squadfillhl~\faQuestionCircle ~\cite{wu2020evaluation} & \moon{15} \\ \hline
\multirow{2}{*}{\begin{tabular}[c]{@{}c@{}}Clinical Decision \\ Support\end{tabular}} & \multicolumn{2}{c}{\begin{tabular}[c]{@{}c@{}}Clinical Summary \\ Generation \end{tabular}} & \begin{tabular}[c]{@{}c@{}}\squadfill~\faFile*~\faQuestionCircle ~\cite{araujo2020adversarial}\\ \squad~\squadfill~\faFile*~\faQuestionCircle ~\cite{wang2020utilizing,fatehi2022towards}\end{tabular} & \squadfill~\faDatabase ~\cite{das2024exposing} & \squad~\faRobot ~\cite{das2024exposing} & \begin{tabular}[c]{@{}c@{}}\squadfill~\faQuestionCircle ~\cite{mireshghallah2022quantifying}\\ \squad~\squadfill~\faQuestionCircle ~\cite{jagannatha2021membership}\\ \squadfillhl~\squadfill~\faQuestionCircle~\cite{hu2022m}\end{tabular} & \squadfillhl~\faQuestionCircle ~\cite{nakamura2020kart} & \squadfillhl~\faQuestionCircle ~\cite{lehman2021does} & \moon{15} \\ \cline{2-10}
 & \multicolumn{2}{c}{\begin{tabular}[c]{@{}c@{}}Automated \\ Clinical Coding\end{tabular}} & \squad~\faFile*~\cite{raja2020adversarial} & \faQuestion & \faQuestion & \squadfillhl~\faQuestionCircle ~\cite{sarkar2024identification} & \moon{8} & \moon{8} & \faQuestion \\ \hline
\multirow{2}{*}{Therapeutics} & \multicolumn{2}{c}{Surgical Treatment} & \faQuestion & \faQuestion & \faQuestion & \moon{1} & \moon{1} & \moon{1} & \moon{8} \\ \cline{2-10}
 & \multicolumn{2}{c}{\begin{tabular}[c]{@{}c@{}}Therapeutic  Prognosis\end{tabular}} & \begin{tabular}[c]{@{}c@{}}\squadfill~\faFile*~\faQuestionCircle ~\cite{mondal2021bbaeg}\\ \squadfill~\faFile*~\faQuestionCircle~\faCommentMedical ~\cite{chai2023additive}\end{tabular} & \squad~\squadfill~\faDatabase~\cite{jagielski2018manipulating} & \moon{8} & \moon{15} & \squadfill~\faQuestionCircle ~\cite{fredrikson2014privacy} & \faQuestion & \squad~\squadfill~\faDatabase~\cite{jagielski2018manipulating} \\ \hline
\multirow{2}{*}{Population Health} & \multicolumn{2}{c}{\begin{tabular}[c]{@{}c@{}}Drug \\ Development Research\end{tabular}} & \moon{8} & \moon{8} & \faQuestion & \moon{1} & \moon{1} & \moon{1} & \faDatabase ~\cite{saini2022predatory} \\ \cline{2-10}
 & \multicolumn{2}{c}{Epidemiology Analysis} & \squad~\faFile*~\cite{meiseles2023vulnerability} & \moon{8} & \faQuestion & \moon{8} & \faQuestion & \faQuestion & \moon{8} \\ \hline
\multirow{2}{*}{\begin{tabular}[c]{@{}c@{}}Patient Health \\ Monitoring\end{tabular}} & \multicolumn{2}{c}{\begin{tabular}[c]{@{}c@{}}Disease Risk \\ Prediction\end{tabular}} & \squadfill~\faFile*~\faQuestionCircle ~\cite{ye2022medattacker, karim2022adversary} & \faQuestion & \faQuestion & \faQuestion & \faQuestion & \faQuestion & \faQuestion \\ \cline{2-10}
 & \multicolumn{2}{c}{\begin{tabular}[c]{@{}c@{}}Monitoring~\& \\ Intervention\end{tabular}} & \squad~\squadfill~\faDatabase~\faFile*~\faQuestionCircle ~\cite{newaz2020adversarial} & \begin{tabular}[c]{@{}c@{}}\squad~\faDatabase ~\cite{shahid2022label}\\ \squad~\squadfill~\faDatabase~\faFile*~\cite{newaz2020adversarial}\\ \squadfill~\faDatabase ~\cite{sun2021data}\end{tabular} & \moon{8} & \faQuestion & \faQuestion & \faQuestion & \moon{15} \\
\specialrule{.2em}{.1em}{.1em}
\end{tabular}
\end{adjustbox}

    \begin{tablenotes}[flushleft]
    \footnotesize
    \item \squad~ white-box knowledge, \squadfillhl~ gray-box knowledge, \squadfill~ black-box knowledge, \faDatabase~ training data control, \faFile*~ testing data control, \faRobot~ model control, \faQuestionCircle~ query access, \faCommentMedical~ explanation access, \moon{15} attack feasible but not done in prior work, \moon{8} attack possible but not done in prior work, \moon{1} attack with impractical or unrealistic threat model, \faQuestion~ not enough information to conclude
    \end{tablenotes}
\end{minipage}%
}
\end{table}

\begin{table}[t]
\caption{Defense research on healthcare AI.}
\label{tab:systemization_all_defense}
\normalsize
\setlength{\tabcolsep}{5pt}
\makebox[\columnwidth][c]{%
\begin{tabular}{lcccclll}
\specialrule{.2em}{.1em}{.1em}
\textbf{} & \multicolumn{4}{c}{\textbf{Defense Type}} &  &  &  \\ \cline{2-5}
\textbf{ID} & Int. & Conf. & Avail. & Exp. & \textbf{Data Modality} & \textbf{Healthcare Domain} & \textbf{Ref.} \\ \specialrule{.2em}{.1em}{.1em}
\multicolumn{1}{l|}{\textbf{D1}} & $\bullet$ & \textbf{-} & \textbf{-} & \multicolumn{1}{c|}{\textbf{-}} & \multicolumn{1}{l|}{\faDna} & \multicolumn{1}{l|}{Multi-omics Diagnostics} & \cite{jagielski2018manipulating} \\
\multicolumn{1}{l|}{\textbf{D2}} & $\bullet$ & \textbf{-} &  \textbf{-} & \multicolumn{1}{c|}{\textbf{-}} & \multicolumn{1}{l|}{\faDna} & \multicolumn{1}{l|}{Multi-omics Diagnostics} & \cite{sarkar2021trapdoor} \\
\multicolumn{1}{l|}{\textbf{D3}} & $\bullet$ & \textbf{-} & \textbf{-} & \multicolumn{1}{c|}{\textbf{-}} & \multicolumn{1}{l|}{\faLungs} & \multicolumn{1}{l|}{Medical Image Detection} & \cite{xiang2023toward} \\
\multicolumn{1}{l|}{\textbf{D4}} & $\bullet$ & \textbf{-} & \textbf{-} & \multicolumn{1}{c|}{\textbf{-}} & \multicolumn{1}{l|}{\faBrain} & \multicolumn{1}{l|}{Medical Image Segmentation} & \cite{zhang2024robust} \\
\multicolumn{1}{l|}{\textbf{D5}} & $\bullet$ & \textbf{-} & \textbf{-} & \multicolumn{1}{c|}{\textbf{-}} & \multicolumn{1}{l|}{\faHeartbeat} & \multicolumn{1}{l|}{Clinical Electrophysiological Diagnosis} & \cite{shao2022defending} \\
\multicolumn{1}{l|}{\textbf{D6}} & $\bullet$ & \textbf{-} & \textbf{-} & \multicolumn{1}{c|}{$\bullet$} & \multicolumn{1}{l|}{\faXRay~\faLaptopMedical} & \multicolumn{1}{l|}{Cross-modal Diagnostics} & \cite{watson2021attack} \\
\multicolumn{1}{l|}{\textbf{D7}} & $\bullet$ & \textbf{-} & $\bullet$ & \multicolumn{1}{c|}{\textbf{-}} & \multicolumn{1}{l|}{\faEye} & \multicolumn{1}{l|}{Medical Image Segmentation} & \cite{zhang2023domain} \\
\multicolumn{1}{l|}{\textbf{D8}} & $\bullet$ & $\bullet$ & \textbf{-} & \multicolumn{1}{c|}{$\bullet$} & \multicolumn{1}{l|}{Network Traffic} & \multicolumn{1}{l|}{Monitoring \& Intervention} & \cite{si2024explainable} \\
\multicolumn{1}{l|}{\textbf{D9}} & $\bullet$ & \textbf{-} & $\bullet$ & \multicolumn{1}{c|}{\textbf{-}} & \multicolumn{1}{l|}{\faXRay~\faAllergies} & \multicolumn{1}{l|}{Medical Image Classification} & \cite{xu2022medrdf} \\
\multicolumn{1}{l|}{\textbf{D10}} & $\bullet$ & \textbf{-} & \textbf{-} & \multicolumn{1}{c|}{\textbf{-}} & \multicolumn{1}{l|}{\faBrain~\faXRay~\faFirstOrder*} & \multicolumn{1}{l|}{Medical Image Classification} & \cite{yang2022defense} \\
\multicolumn{1}{l|}{\textbf{D11}} & $\bullet$ & \textbf{-} & \textbf{-} & \multicolumn{1}{c|}{\textbf{-}} & \multicolumn{1}{l|}{\faEye} & \multicolumn{1}{l|}{Medical Image Classification} & \cite{lal2021adversarial} \\
\multicolumn{1}{l|}{\textbf{D12}} & $\bullet$ & \textbf{-} & \textbf{-} & \multicolumn{1}{c|}{\textbf{-}} & \multicolumn{1}{l|}{\faXRay~\faAllergies} & \multicolumn{1}{l|}{Diagnostic Imaging} & \cite{jin2023backdoor} \\
\multicolumn{1}{l|}{\textbf{D13}} & $\bullet$ & \textbf{-} & \textbf{-} & \multicolumn{1}{c|}{\textbf{-}} & \multicolumn{1}{l|}{\faLungs~\faBrain~\faDiagnoses} & \multicolumn{1}{l|}{Medical Image Detection} & \cite{joel2021adversarial} \\
\multicolumn{1}{l|}{\textbf{D14}} & $\bullet$ & \textbf{-} & \textbf{-} & \multicolumn{1}{c|}{\textbf{-}} & \multicolumn{1}{l|}{\faFileWord} & \multicolumn{1}{l|}{Clinical Summary Generation} & \cite{wang2020utilizing} \\
\multicolumn{1}{l|}{\textbf{D15}} & \textbf{-} & $\bullet$ & \textbf{-} & \multicolumn{1}{c|}{\textbf{-}} & \multicolumn{1}{l|}{\faDna} & \multicolumn{1}{l|}{Multi-omics Diagnostics} & \cite{backes2017identifying} \\
\multicolumn{1}{l|}{\textbf{D16}} & \textbf{-} & $\bullet$ & \textbf{-} & \multicolumn{1}{c|}{\textbf{-}} & \multicolumn{1}{l|}{Motion Sensor} & \multicolumn{1}{l|}{Monitoring \& Intervention} & \cite{boutet2021dysan} \\
\multicolumn{1}{l|}{\textbf{D17}} & \textbf{-} & $\bullet$ & \textbf{-} & \multicolumn{1}{c|}{\textbf{-}} & \multicolumn{1}{l|}{\faHeartbeat} & \multicolumn{1}{l|}{Clinical Electrophysiological Diagnosis} & \cite{abuadbba2020can} \\
\multicolumn{1}{l|}{\textbf{D18}} & \textbf{-} & $\bullet$ & \textbf{-} & \multicolumn{1}{c|}{\textbf{-}} & \multicolumn{1}{l|}{Body Measurements} & \multicolumn{1}{l|}{Therapeutic Prognosis} & \cite{xue2016privacy} \\
\multicolumn{1}{l|}{\textbf{D19}} & \textbf{-} & $\bullet$ & \textbf{-} & \multicolumn{1}{c|}{\textbf{-}} & \multicolumn{1}{l|}{Contact Patterns} & \multicolumn{1}{l|}{Epidemiology Analysis} & \cite{romijnders2024protect} \\
\multicolumn{1}{l|}{\textbf{D20}} & \textbf{-} & $\bullet$ & \textbf{-} & \multicolumn{1}{c|}{\textbf{-}} & \multicolumn{1}{l|}{\faFileWord~\faLaptopMedical} & \multicolumn{1}{l|}{Cross-modal Diagnostics} & \cite{xu2020federated} \\
\multicolumn{1}{l|}{\textbf{D21}} & \textbf{-} & $\bullet$ & \textbf{-} & \multicolumn{1}{c|}{\textbf{-}} & \multicolumn{1}{l|}{\faFileImage} & \multicolumn{1}{l|}{Medical Image Classification} & \cite{tian2022confoundergan} \\
\multicolumn{1}{l|}{\textbf{D22}} & \textbf{-} & $\bullet$ & \textbf{-} & \multicolumn{1}{c|}{\textbf{-}} & \multicolumn{1}{l|}{\faLaptopMedical} & \multicolumn{1}{l|}{Therapeutic Effect Prediction} & \cite{das2023twin} \\
\multicolumn{1}{l|}{\textbf{D23}} & \textbf{-} & $\bullet$ & \textbf{-} & \multicolumn{1}{c|}{\textbf{-}} & \multicolumn{1}{l|}{\faXRay~\faFirstOrder*} & \multicolumn{1}{l|}{Medical Image Classification} & \cite{zhou2023unified} \\
\multicolumn{1}{l|}{\textbf{D24}} & \textbf{-} & $\bullet$ & \textbf{-} & \multicolumn{1}{c|}{\textbf{-}} & \multicolumn{1}{l|}{\faEye} & \multicolumn{1}{l|}{Medical Image Classification} & \cite{paul2021defending} \\
\multicolumn{1}{l|}{\textbf{D25}} & \textbf{-} & $\bullet$ & \textbf{-} & \multicolumn{1}{c|}{\textbf{-}} & \multicolumn{1}{l|}{\faXRay} & \multicolumn{1}{l|}{Diagnostic Imaging} & \cite{yala2021neuracrypt} \\
\multicolumn{1}{l|}{\textbf{D26}} & \textbf{-} & $\bullet$ & \textbf{-} & \multicolumn{1}{c|}{\textbf{-}} & \multicolumn{1}{l|}{\faLaptopMedical} & \multicolumn{1}{l|}{Cross-modal Diagnostics} & \cite{chou2018fully} \\
\multicolumn{1}{l|}{\textbf{D27}} & \textbf{-} & $\bullet$ & \textbf{-} & \multicolumn{1}{c|}{\textbf{-}} & \multicolumn{1}{l|}{\faLaptopMedical} & \multicolumn{1}{l|}{Cross-modal Diagnostics} & \cite{rohanian2022privacy} \\
\multicolumn{1}{l|}{\textbf{D28}} & \textbf{-} & $\bullet$ & \textbf{-} & \multicolumn{1}{c|}{\textbf{-}} & \multicolumn{1}{l|}{\faFileImage} & \multicolumn{1}{l|}{Medical Image Classification} & \cite{sun2024medical} \\
\multicolumn{1}{l|}{\textbf{D29}} & \textbf{-} & $\bullet$ & \textbf{-} & \multicolumn{1}{c|}{\textbf{-}} & \multicolumn{1}{l|}{\faDna} & \multicolumn{1}{l|}{Multi-omics Diagnostics} & \cite{chen2020differential} \\
\multicolumn{1}{l|}{\textbf{D30}} & \textbf{-} & \textbf{-} & $\bullet$ & \multicolumn{1}{c|}{\textbf{-}} & \multicolumn{1}{l|}{\faDna} & \multicolumn{1}{l|}{Multi-omics Diagnostics} & \cite{zheng2023multi} \\
\multicolumn{1}{l|}{\textbf{D31}} & \textbf{-} & \textbf{-} & $\bullet$ & \multicolumn{1}{c|}{\textbf{-}} & \multicolumn{1}{l|}{\faFileImage~\faXRay~\faDiagnoses} & \multicolumn{1}{l|}{Medical Image Segmentation} & \cite{lin2024safeguarding} \\
\multicolumn{1}{l|}{\textbf{D32}} & \textbf{-} & \textbf{-} & \textbf{-} & \multicolumn{1}{c|}{$\bullet$} & \multicolumn{1}{l|}{\faBrain} & \multicolumn{1}{l|}{Medical Image Classification} & \cite{sihag2024explainable} \\
\multicolumn{1}{l|}{\textbf{D33}} & \textbf{-} & \textbf{-} & \textbf{-} & \multicolumn{1}{c|}{$\bullet$} & \multicolumn{1}{l|}{\faHeartbeat} & \multicolumn{1}{l|}{Clinical Electrophysiological Diagnosis} & \cite{maurer2024explainable} \\
\multicolumn{1}{l|}{\textbf{D34}} & \textbf{-} & \textbf{-} & \textbf{-} & \multicolumn{1}{c|}{$\bullet$} & \multicolumn{1}{l|}{\faLungs~\faXRay~\faShoePrints} & \multicolumn{1}{l|}{Medical Image Classification} & \cite{han2021advancing} \\
\multicolumn{1}{l|}{\textbf{D35}} & \textbf{-} & \textbf{-} & \textbf{-} & \multicolumn{1}{c|}{$\bullet$} & \multicolumn{1}{l|}{\faFileImage} & \multicolumn{1}{l|}{Medical Image Segmentation} & \cite{cortacero2023evolutionary} \\
\specialrule{.2em}{.1em}{.1em}
\end{tabular}%
}
    \begin{tablenotes}[flushleft]
    \scriptsize
    \item \faDna~Genomic Data, \faLungs~Lung CT, \faBrain~Brain MRI, \faShoePrints~Knee MRI, \faHeartbeat~Electrogram, \faXRay~Chest X-Ray, \faLaptopMedical~EHR, \faEye~Fundus Image, \faAllergies~Dermoscopic Image, \faFirstOrder*~Histological Image, \faDiagnoses~Mammogram, \faFileWord~Text Data, \faFileImage~Image Data, Int. = Integrity, Conf. = Confidentiality, Avail. = Availability, Exp. = Explainability
    \end{tablenotes}
\end{table}

\subsection{Diagnostic Imaging}
\label{subsec:image}

\bsub{Domain Overview.}
Healthcare AI for diagnostic imaging performs tasks in various fields of medicine, and is often categorized into \textit{classification}, \textit{detection}, and \textit{segmentation} tasks in biomedical literature. In healthcare, classification tasks separate medical images into categories according to the type of image or the presence of different conditions for a specific disease~\cite{groh2021evaluating}.
Detection tasks involve establishing spatial localization of regions of interest in medical images~\cite{schmitt2024machine,lee2019explainable}.
Finally, segmentation tasks focus on grouping at the pixel or voxel level for a given image type~\cite{wang2021annotation}.

\bsub{Domain Particularities and Practical Challenges.}
One challenge with defending diagnostic imaging AI is the involvement of multiple image modalities (e.g., CT, MRI, X-ray), resolutions (nanometers in microscopy to centimeters in tissue samples), scanner hardware, digitization protocols, and demographic-specific anatomy during diagnostic procedures~\cite{yu2023xcheck,zhang2023domain,cortacero2023evolutionary}. Perturbations and defenses that work in one domain may not generalize well to another due to domain-specific image statistics (e.g., fundus vs. chest X-ray textures)~\cite{bortsova2021adversarial}. Even within the same modality, healthcare images often represent diagnostically critical areas in very small regions (e.g., lesions, tumors, nodules, organ boundaries, etc.)~\cite{sun2024medical}. Since triggers that visibly distort medical features will likely be flagged by human experts, prior work has attempted to insert perturbations more stealthily, such as in uncritical areas~\cite{nwadike2020explainability} or frequency-domain perturbations~\cite{feng2022fiba}. Even so, medical images often have consistent spatial homogeneity and structure, and adversarial attacks may create perturbations that are statistically dissimilar to real medical images~\cite{ma2021understanding}.

\isub{Unique Aspects in Classification Task:}
Healthcare images often have very subtle inter-class differences, making it easier for adversarial noise to shift predictions~\cite{ma2021understanding}. For instance, tumor vs. no tumor hinges on very fine pixel patterns, unlike dogs vs. cats in natural images. To address this, Xu et al.~\cite{xu2022medrdf} proposed MedRDF, which predicts diagnostic results based on majority voting, as traditional preprocessing-based defenses, such as random resizing or compression, may remove small but critical diagnostic features. Furthermore, healthcare AIs often learn from limited and highly specialized datasets~\cite{mozaffari2014systematic} and are rarely updated due to models existing in regulated, static hospital systems, making memorization more likely~\cite{wu2020evaluation}. This has led many medical models (such as MedCLIP) to rely on unpaired or weakly aligned data to overcome the shortage of labeled clinical datasets, making them intrinsically ``poisoned''~\cite{jinbackdoor}.

\isub{Unique Aspects in Detection Task:}
Since existing adversarial defenses do not account for semantic interdependence among landmarks in healthcare image detection tasks, Yao et al.~\cite{yao2020miss} were able to control the output for selected anatomical landmarks while leaving others unchanged. However, unlike object detection in traditional ML tasks, altering just a few landmark positions on a medical image (such as in cephalometric analysis used in orthodontics and surgery planning) may affect anatomical dependencies between landmarks, which may be used to detect perturbations or to reinforce model constraints.

\isub{Unique Aspects in Segmentation Task:}
Unlike adversarial examples in classification tasks that target a single class, adversarial targets in medical image segmentation are masks and involve optimizing many individual pixels in the generation process~\cite{ozbulak2019impact}. Traditional unlearnable examples that work for classification and detection fail to disrupt training in segmentation tasks because they do not interfere with clinically critical visual features like contours and textures~\cite{lin2024safeguarding}. However, such spatial features may also cause healthcare segmentation models to be more vulnerable to inference attacks when spatial consistency is leveraged~\cite{chobola2023membership}.

\clearpage
\begin{tcolorbox}[compactbox]
\textbf{Research Opportunities.}
We found that prior research on diagnostic imaging systems prefers certain data modalities (such as X-ray images) over others. One key implication for future research observed from this challenge is the development of suitable foundation models that can generalize within and across modalities and datasets, which has been started in recent years~\cite{rood2024toward,wang2022medclip, zhang2023biomedclip}.
In addition, with the rise of generative AI technologies, large models such as vision-language models (VLMs)~\cite{pan2025medvlm} and diffusion models have also been applied to diagnostic tasks~\cite{lin2024stable}. Although these models achieve strong performance on medical diagnostic tasks, they remain vulnerable to adversarial attacks (e.g., evasion attacks) that can induce unsafe or incorrect outputs. In \S\ref{subsec:vlm-attack}, we use medical VLMs as a representative example and empirically demonstrate their vulnerability to evasion attacks.
On the other hand, generative AI has also been leveraged to produce adversarial examples~\cite{zhou2021machine}. How to defend against those while utilizing them for healthcare diagnostics, such as to generate synthetic data to mitigate dataset biases~\cite{varoquaux2022machine}, is another direction worth researching.
\end{tcolorbox}

\subsection{Clinical Electrophysiological Diagnosis}
\label{subsec:electrogram}

\bsub{Domain Overview.}
Electrograms are used to record and visualize electrical activity within the body. ML methods have been used to diagnose arrhythmias~\cite{hannun2019cardiologist,clifford2017af} with electrocardiograms (ECG), schizophrenia with electroencephalograms (EEG)~\cite{shim2016machine}, neuromuscular disorders~\cite{subasi2013classification} with electromyograms (EMG), and to classify sleep stages~\cite{rahman2018sleep} with electrooculograms (EOG). Notably, Kiranyaz et al.~\cite{kiranyaz2015real} were the first to use DNNs over 1D signals, particularly for ECG classification tasks.

\bsub{Domain Particularities and Practical Challenges.}
Unlike images, healthcare AI using electrograms as inputs faces challenges in handling multivariate time series data. Abuadbba et al.~\cite{abuadbba2020can} found that split learning with 1D CNNs trained on medical waveforms has increased risks of privacy leakage due to the preservation of time-dependent features in intermediate activations.
Adversarial electrogram examples also demand both temporal coherence and physiological realism. Plus, real-world recording devices include filtering to combat noise (e.g., notch filter for powerline interference, low-pass filter for muscle noise, etc.).
For instance, ECGs are expected to follow precise P-QRS-T patterns and must respect beat-level timing~\cite{ono2021application}. Existing image-targeted attacks are not applicable since $L_p$ norm-based constraints produce square-wave patterns that stand out in ECG curves. Therefore, Chen et al.~\cite{chen2020ecgadv} generated imperceptible perturbations by quantifying perceptual similarities of line curves, successfully evading both arrhythmia classification systems and cardiology experts. They showed that while electrograms are structured, even trained cardiologists may fail to distinguish perturbations targeting valleys, as diagnoses tend to focus on the peaks of ECG waves~\cite{chen2020ecgadv}.

\begin{tcolorbox}[compactbox]
\textbf{Research Opportunities.}
Electrogram waveforms' regularity and known medical interpretation standards (e.g., R-peak detection, rhythm morphology) make them suitable for defense methods that smooth or regularize model learning around core physiological features~\cite{shao2022defending}. In addition, the temporally dynamic nature makes it difficult for explainable models to consistently interpret features across cases~\cite{maurer2024explainable}, warranting further research.
Meanwhile, with the rapid growth and widespread adoption of edge AI devices, electrogram signals are increasingly leveraged in wearable and edge medical systems for continuous health monitoring and user authentication~\cite{arias2021inexpensive, luo2020oculock, shao2025know}. However, this trend also expands the attack surface, as such signals can be exploited for both security attacks (e.g., spoofing against biosignal-based authentication) and privacy attacks (e.g., sensitive health inference and identity exposure)~\cite{eberz2017broken, krish2024biosignal, martinovic2012feasibility, ye2025bpsniff}. Therefore, understanding the security and privacy risks of electrogram signals remains a major challenge and an important future research direction for the safe deployment of medical devices.

\end{tcolorbox}

\subsection{Multi-omics Diagnostics}
\label{subsec:multiomics}

\bsub{Domain Overview.}
Multi-omics diagnostics analyzes data from various omics sources and data acquisition technologies, including genomics, epigenomics, metabolomics, and proteomics~\cite{zheng2023multi}. When considering early cancer detection, multi-omics data can include information about mutations, gene expression, and copy number variation~\cite{schulte2021integration}.
Meanwhile, multi-omics approaches can also be used to more accurately classify chronic kidney disease. By operating ML on molecular data composed of kidney biopsy, blood, and urine samples, Eddy et al.~\cite{eddy2020integrated} classified patients into molecularly defined subgroups that better reflect the underlying mechanisms associated with the disease.

\bsub{Domain Particularities and Practical Challenges.}
Unlike image and waveform data, genomic sequences often involve thousands of input features (e.g., SNPs, gene expressions, etc.), and the model must generalize from sparse, structured data~\cite{skovorodnikov2024fimba}. In addition, pathology models often work with gigapixel histopathological slides and rely on slide-level (rather than pixel-level) labels~\cite{ghaffari2022adversarial}.
The multi-source and interwoven nature of omics data leads to several concerns. First, family-level dependencies are strong, causing health data from parents to expose children's traits and vice versa~\cite{berrang2018dissecting}. Second, biological datasets often contain latent demographic biases (e.g., race, gender); a conclusion stemming from a biased dataset not only affects an individual but also the entire group sharing the genetic traits~\cite{sarkar2021trapdoor}. In addition, given the generally limited genomic sample size and large feature space, overfitting in models like CNNs and Lasso regression elevates their susceptibility to privacy attacks~\cite{chen2020differential}. Finally, epigenetic data bridge phenotype and genotype in ways that general data modalities do not. For instance, methylation data can reveal both inherited traits and environmental exposures~\cite{hagestedt2020membership, backes2017identifying}.

\begin{tcolorbox}[compactbox]
\textbf{Research Opportunities.}
Compared to the number of market-cleared healthcare AI by the FDA in radiology (521; 76\%) and cardiology (102; 10\%)~\cite{Fornell_2024}, multi-omics diagnostics models have received less market and research attention due to data scarcity. Though open platforms for citizen science or ancestry exploration facilitate the collection of genetic data from the public, the integrity and confidentiality of these datasets are a large concern and offer opportunities for future research. Not only can adversaries inject manipulated samples into large collaborative studies~\cite{montserrat2023adversarial}, but data handling during ownership transition can be controversial, as in the case of the recent 23andMe bankruptcy~\cite{Duffy_2025}. Recent work has started to explore genomic beacons as a method of privacy-preserving data sharing~\cite{hagestedt2019mbeacon}, though they were also shown to have vulnerability against membership inference~\cite{shringarpure2015privacy}, and even genome reconstruction attacks~\cite{ayoz2021genome}.
\end{tcolorbox}

\subsection{Cross-modal Diagnostics}
\label{subsec:crossmodal}

\bsub{Domain Overview.}
Traditional medical expert systems often employ feature-level fusion or rule-based reasoning, with performance significantly affected by subjective expert-defined rules that cannot be dynamically updated. Especially for multi-source, multi-modal healthcare data, these fall short of offering integration, reasoning, and interactive decision support.
Meanwhile, cross-modal AI employs a multi-network link or network reconstruction based on DNN with feature coupling~\cite{lyu2022multimodal}, extracting high-level features from the original data's bottom-up representation. This helps achieve intelligent auxiliary support in complex in-hospital scenes. Often, entity mining is used to perform semantic perception and correlation mining to support data fusion~\cite{kavakiotis2017machine}. More recent works further leverage LLMs to handle complex multimodal medical data, including clinical text and multimodal EHR streams, often through text serialization or retrieval-augmented fusion pipelines~\cite{yang2022large, lee2025clinical}.

\bsub{Domain Particularities and Practical Challenges.}
EHRs are commonly used in healthcare to store various data modalities to support the efficient retrieval of patient information. Unlike continuous sensor data or structured tabular data, EHRs involve frequently recurring events (like vital signs), episodic events, and sometimes missing data~\cite{xu2020federated, sun2018identify}. Adversarial examples must respect this temporal granularity to be effective and realistic~\cite{joe2021machine}. EHR-based predictive models are particularly sensitive to longitudinal perturbations~\cite{an2019longitudinal}, and prior work has shown the possibility to leverage explainability (e.g., SHAP scores) as an anomaly detection feature to identify attacks that distort these semantic patterns~\cite{watson2021attack}.

\begin{tcolorbox}[compactbox]
\textbf{Research Opportunities.}
Contrasting with more general-purpose datasets like ImageNet, EHR datasets contain clinically grounded feature semantics. The use of explainability metrics to identify adversarial examples forces attackers to create semantically consistent perturbations. However, semantically coherent adversarial inputs have already been demonstrated in emerging LLM systems through methods such as GCG~\cite{zou2023universal} and AutoDAN~\cite{liuautodan}. These attacks can remain fluent and locally consistent while still steering model behavior, posing a significant challenge to LLM-based multimodal healthcare analysis pipelines.
Meanwhile, since patients from different hospitals may have different clinical characteristics, similarity-preserving and privacy-preserving data retrieval across institutions for cohort construction can be challenging~\cite{xu2020federated}.
In this case, FL systems can help facilitate global model updates without the need for institutions to share raw patient data.

\end{tcolorbox}

\section{Clinical Decision Support} \label{sec:Clinical_Decision_Support}

\bsub{Domain Overview.}
The emergence of large language models (LLMs) has expanded clinical decision support beyond traditional rule-based systems and conventional clinical NLP pipelines. In addition to administrative tasks that support clinical decisions, such as generating discharge summaries~\cite{arora2023promise} and assigning ICD-10 codes to healthcare records~\cite{ji2022unified}, LLMs are increasingly explored as clinical copilots for medication review, chart abstraction, evidence summarization, and recommendation drafting under clinician supervision~\cite{ong2025large}. Meanwhile, emerging agentic AI systems further extend these capabilities by combining LLM reasoning with planning and tool use (e.g., guideline lookup and clinical calculators), enabling multi-step workflows that more closely resemble clinical decision-support processes~\cite{goodell2025large, liu2026benchmarking}.

\bsub{Domain Particularities and Practical Challenges.}
Clinical texts often contain domain-specific jargon, frequent typos and misspellings, and abbreviations of medical terms, making it hard to distinguish natural anomalies from adversarial perturbations~\cite{wang2020utilizing, patel2023chatgpt}. Exploiting this, prior work has considered edit adversaries in clinical settings, where adversaries used naturally occurring spelling errors and medical synonym substitutions to drop the performance of biomedical named-entity recognition tasks~\cite{araujo2020adversarial}, as well as leveraged imperceptible typo-based attacks to downgrade the performance of clinical ICD-code prediction systems~\cite{raja2020adversarial}.
For LLM-based and especially agentic clinical decision-support systems, the attack surface further extends beyond textual perturbations. The adversaries may exploit retrieval, tool selection, intermediate reasoning, and even tool outputs to conduct attacks, including indirect prompt injection through retrieved documents or tool-returned content~\cite{zhan2024injecagent, greshake2023not}.

Meanwhile, because biomedical language models are often trained on sparse or rare event data, smaller model sizes are common, leading to higher overfitting risks~\cite{jagannatha2021membership}. Clinical text also contains repetitive patterns (e.g., diagnosis templates, drug interactions) that increase the risk of memorization~\cite{mireshghallah2022quantifying}. Sarkar et al.~\cite{sarkar2024identification} have explored the possibility of membership inference attacks against ICD-coding classification and found that even de-identified clinical notes are not sufficient to guarantee patient privacy.

\begin{tcolorbox}[compactbox]
\textbf{Research Opportunities.}
The inherently messy nature of clinical texts leads to several opportunities. While it may be easier to hide adversarial perturbations among typos, the impact of various denoising strategies on adversarial manipulations, deployed in the context of healthcare language models, needs to be thoroughly evaluated. In addition, the unstructured clinical summaries are often coupled with structured features (e.g., vitals, medications) in the EHR. This multi-modality is underused in prior work but uniquely valuable for cross-validating inconsistencies in clinical applications~\cite{wang2020utilizing}. For emerging medical agentic AI systems, indirect prompt injection poses a serious threat to functional integrity because it can corrupt the agent's decision pipeline and induce unsafe actions. We empirically demonstrate this attack capability in \S \ref{subsec:prompt-inject}. Existing defenses include model-level methods that improve intrinsic robustness~\cite{chen2025struq, chen2025meta} and system-level methods that enforce runtime security policies~\cite{li2025drift, shi2025progent}. However, these defenses still degrade in dynamic real-world environments, where tools, context, and attack strategies continually change~\cite{li2026agentdyn}, highlighting the need for stronger medical-agent-specific protections.
\end{tcolorbox}

\section{Therapeutics}
\label{sec:Therapeutics}

\bsub{Domain Overview.}
Besides diagnostics and administrative tasks, AI in \textit{surgical treatments} such as intraoperative guidance~\cite{sganga2019offsetnet} and surgical robots~\cite{kurmann2017simultaneous,marban2018estimation} provides enhanced visualization and localization. Moreover, the emerging use of generative AI in healthcare has enabled the localization of a bronchoscope in the lung even in less conserved regions~\cite{sganga2019offsetnet}.
For \textit{treatment response prediction}, AI can assist in estimating treatment effects, helping patients and clinicians evaluate the effectiveness of current treatment plans~\cite{zhou2017causal}.

\bsub{Domain Particularities and Practical Challenges.}
Currently, no prior work has attacked AI for intraoperative guidance or surgical robots. While performing attacks is possible, proprietary surgical treatment AI may not be easily accessible by attackers or researchers. Surgical AI may be similar in input to diagnostic imaging in addition to positional data, yet even if access is granted or assuming black-box attacks, attackers need to consider the real-world realizability of adversarial input, especially for surgical robots that take in real-time inputs. Profit may be minimal compared to the attacker's cost of sabotaging the system.

For treatment response prediction, besides taking EHR as input and needing to adapt to its particularities~\cite{mondal2021bbaeg}, defenses should not affect the system's real-time responsiveness. For instance, Fredrikson et al.~\cite{fredrikson2014privacy} and Xue et al.~\cite{xue2016privacy} found that privacy-preserving computations, such as differential privacy, may degrade model accuracy and limit model interactivity and adaptability.

\begin{tcolorbox}[compactbox]
\textbf{Research Opportunities.}
Few prior works have explored AI systems for treatment response prediction. In addition to the real-world and real-time requirements, treatment-response models are often not trusted by physicians for clinical deployment~\cite{vanRoyenLeakyPipeline}. Fortunately, digital twin technology has emerged as a promising solution for computer-assisted surgery and prognosis for its ability to provide high-granularity feedback that reflects patient characteristics in real time~\cite{asciak2025digital,das2023twin}, presenting opportunities for S\&P research in this area.
\end{tcolorbox}

\section{Population Health}
\label{sec:Population_Health}

\bsub{Domain Overview.}
In addition to individual patient care, healthcare AI has contributed to population-level applications. This is evident in \textit{drug development research}, including target identification~\cite{mountjoy2021open}, drug discovery~\cite{segler2018planning}, and drug-drug interaction (DDI)~\cite{baxter2010stockley}, where the abundant availability of drug-related information from biomedical texts, EHRs, and public databases provides fertile ground for literature-based extractions~\cite{hong2020novel}.
In addition, predictive models based on chemical and biological knowledge are useful in treating DDI as a link prediction task to detect the presence or absence of drug-pair interactions~\cite{yang2019white}.
On the other hand, healthcare AI can help with \textit{epidemiology analysis}~\cite{brownstein2023advances}.
AI used for pandemic early warning, such as HealthMap~\cite{freifeld2008healthmap}, can compile web texts for indications of infectious disease occurrences.
The data gathered can also be used for real-time outbreak analysis~\cite{bhatia2021using} or integrated into healthcare IoT devices for real-time infection identification~\cite{alavi2022real}. However, all these population-level pipelines increasingly rely on large-scale aggregation and voluntary donation of sensitive health data, making privacy guarantees and user trust important factors in data availability and deployment feasibility~\cite{wang2024role, gaballah2025s}.

\bsub{Domain Particularities and Practical Challenges.}
Predatory research~\cite{richtig2018problems} has become a problem for healthcare AI that explores drug discovery or extracts DDI information from notable biomedical databases such as PubMed and SemMedDB. As predatory content can be semantically and syntactically high-quality, detection requires understanding citation behavior, journal reputation, and semantic content relevance~\cite{saini2022predatory}. On the other hand, epidemiology data, such as contact tracing risk scores, are immediately linked to actionable health decisions (e.g., self-isolation, testing)~\cite{romijnders2024protect}. Meiseles et al.~\cite{meiseles2023vulnerability} also successfully performed evasion attacks on Pangolin, the de facto standard in SARS-CoV-2 lineage tracking. The attack perturbs only 2–3 base pairs in the virus genome, which causes incorrect lineage assignment. Compromising these metrics and tools risks misinformed outbreak responses or policy decisions.
For privacy concerns, recent work shows that de-identification and anonymization can fail in practice, including re-identification from smartwatch health data~\cite{lange2025slice} and leakage from locally recoded $k$-anonymized clinical datasets even without auxiliary data~\cite{chhillar2025exposing}. As medical and public-health workflows increasingly incorporate LLM pipelines, Protected Health Information (PHI) leakage and privacy attacks in medical-domain LLM systems add further risk~\cite{kang2025can}. These privacy risks may also reduce willingness to donate health data, affecting participation and dataset representativeness~\cite{wang2024role, gaballah2025s}.

\begin{tcolorbox}[compactbox]
\textbf{Research Opportunities.}
AI in population health management operates on extremely imbalanced data due to the localization of infected individuals or sampled populations~\cite{leslie2021does,chowkwanyun2020racial,obermeyer2019dissecting}. The issue of fairness and dataset bias is a prominent one. Not only can a few maliciously modified samples disproportionately influence the model's behavior against the underrepresented classes, but overfitting to the training data can also potentially make inference attacks easier. While prior work has proposed some mitigation methods~\cite{chen2023algorithmic,richardson2021framework}, a systematic overview of current strategies, as well as metrics to measure AI fairness, is needed lest an unfair model gets clinically deployed. For privacy concerns, privacy-preserving mechanisms such as secure multi-party computation and federated learning can help reduce direct data exposure. However, we empirically demonstrate in \S\ref{subsec:data-reconstruct} that data reconstruction attacks can still expose private training samples. Therefore, how to securely protect public-health data throughout its collection, sharing, and downstream use remains an open direction.
\end{tcolorbox}

\section{Patient Health Monitoring} \label{sec:Patient_Health_Monitoring}

\bsub{Domain Overview.}
Prior research efforts have been dedicated to \textit{disease risk prediction}, including disease and cancer susceptibility, survivability, occurrence, and recurrence~\cite{ming2020machine,dai2022survival,placido2023deep,kucukkaya2023predicting}, taking into account genetics, inheritance, lifestyle, and other relevant traits of a person. Systems using AI to \textit{monitor} personal lifestyles and provide appropriate \textit{interventions} in non-hospital settings also offer a convenient and less intrusive way for individuals to manage their health~\cite{mondol2016medrem}. Although privacy and real-time responsiveness are significant, prior work in this domain focused primarily on compromising the integrity of system outcomes. Recent works further extend this domain to LLM-based healthcare consultation, where users increasingly ask language models to assess symptoms, explain health conditions, or provide self-care suggestions, raising new concerns about privacy, trust, and appropriate reliance~\cite{liu2025prevalence}.

\bsub{Domain Particularities and Practical Challenges.}
Unlike static data environments, AI in patient health monitoring predicts based on real-time continuous temporal sequences, which adversarial attacks must consistently subvert over time~\cite{newaz2020adversarial}. As static models fail to generalize to dynamic activities, Boutet et al.~\cite{boutet2021dysan} introduced a transfer learning framework that selects optimal sanitization models based on the characteristics of motion sensor data from wearables before they are uploaded or processed. The reliance on wireless data transmission also complicates the application of centralized learning and motivates edge-based approaches. Si et al.~\cite{si2024explainable} apply FL to preserve data locality and combine it with explainable AI to provide interpretable outputs.
Additionally, as LLM-based health consultation becomes prevalent, an additional challenge is ensuring public awareness and acknowledgment of the limitations and imperfections of these models (e.g., uncertainty, hallucinations, and over-reliance), especially when outputs may influence health decisions~\cite{liu2025prevalence}. Another practical challenge is that connected health-monitoring software and devices are not always updated in a timely manner, and older versions may remain vulnerable to newer attacks~\cite{kustosch2025patching}.
Finally, wearable monitoring devices operate in noisy, real-world environments, which may increase attack effectiveness. For instance, due to sensor noise and contextual variability (e.g., walking vs. running on uneven terrain), adversaries can flip activity labels with minimal suspicion~\cite{shahid2022label}.

\begin{tcolorbox}[compactbox]
\textbf{Research Opportunities.}
The persistent flow of voluminous sensitive data due to continuous monitoring calls for inline, low-latency protections. In addition, both adversarial attacks and defenses need to be transferable and usable on different monitoring devices. Recent advances in applying tinyML to healthcare serve as a promising foundation for further explorations~\cite{tsoukas2021review}. For AI involved in critical life monitoring and intervention tasks, besides detecting malicious behavior, it is necessary to enforce prompt system recovery to ensure patient safety~\cite{si2024explainable}. In addition, future work should improve the robustness and reliability of medical LLM systems while also providing clearer public communication and calibrated disclosure of their limitations to support appropriate use. Another important direction is strengthening edge-AI security for health-monitoring deployments, including patch-aware defenses, secure update mechanisms, and protections that remain effective when devices run heterogeneous or outdated software versions.
\end{tcolorbox}

\needspace{6\baselineskip}
\section{Healthcare AI Case Studies}
\label{sec:exp}

\subsection{Adversarial Attacks on Medical VLMs}
\label{subsec:vlm-attack}

\bsub{Overview.} Vision-language models (VLMs) are among the most promising approaches for bridging visual and textual modalities. In the medical domain, they support cross-modal diagnostic tasks by analyzing medical images and generating corresponding clinical outputs. For example, VLMs have been applied to medical visual question answering (VQA), where they answer clinical questions based on X-ray, CT, and MRI images~\cite{pan2025medvlm}. However, as we have discussed in \S\ref{sec:Medical_Diagnostics_Systems}, medical VLMs can be vulnerable to adversarial attacks, for which the attacker adds human-unnoticeable perturbations to medical images and misleads the model into generating wrong outputs. In this case study, we demonstrate the effectiveness of such adversarial attacks and highlight the need for more advanced defense mechanisms.

\begin{figure}[h]
  \centering
  \includegraphics[width=0.7\textwidth]{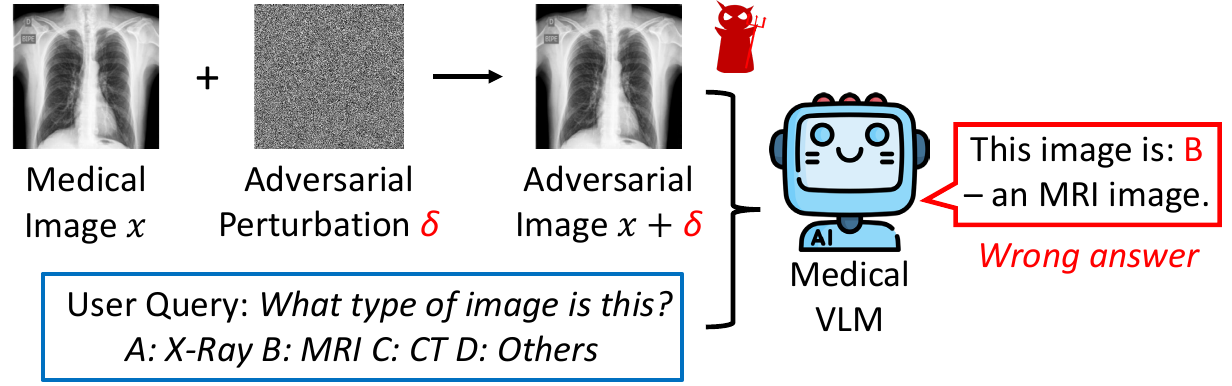}
  \caption{Adversarial examples against the medical VLM.}
  \label{fig:vlm-ae}
\end{figure}

\bsub{Threat Model.} Following the formulation of classic adversarial examples~\cite{goodfellow2014explaining, madry2017towards, carlini2017towards}, we assume that the adversary can add a $p$-norm-bounded perturbation to medical images with the goal of causing the VQA model to produce an incorrect prediction. Regarding the attacker's knowledge, we assume a white-box setting in which the adversary has full access to the model parameters.

\bsub{Attack Method.} We adopt the projected gradient descent (PGD) algorithm to generate targeted adversarial perturbations. Given a clean medical image $x$ and a user query $q$, the adversary seeks a small perturbation $\delta$ such that the perturbed image $x^{\mathrm{adv}} = x + \delta$ remains visually similar to the original image but causes the medical VQA model to output a target incorrect answer. Specifically, the adversary first selects a wrong target answer $y_t$, then back-propagates the loss with respect to $y_t$ to the input image and iteratively updates the image in the gradient direction. After each step, the perturbation is projected back to the allowed $p$-norm ball to ensure that it satisfies the perturbation budget. As illustrated in Figure~\ref{fig:vlm-ae}, the adversary perturbs a lung X-ray image so that, for the question ``What type of image is this?'', the model incorrectly predicts ``MRI'' instead of the correct answer ``X-ray''.

\bsub{Experiment Settings.} We implement the attack on a medical VQA dataset adopted from~\cite{pan2025medvlm}, which contains 900 medical question-answer pairs spanning CT, X-ray, and MRI images. We set the perturbation budget to $\frac{4}{255}$ to ensure that the adversarial images remain visually indistinguishable from the benign ones. We conduct targeted attacks by specifying an incorrect target answer for each question-answer pair. To evaluate attack performance, we use three metrics: benign accuracy, attack accuracy, and targeted attack success rate (ASR).

\bsub{Empirical Results.}
We demonstrate the performance of targeted adversarial attacks against the medical VLM in Table~\ref{tab:vlm-ae}. The results show that the attack is effective across all three imaging modalities. On benign inputs, the medical VLM achieves accuracies of 96.3\%, 72.7\%, and 70.0\% on MRI, CT, and X-ray images, respectively. Under targeted PGD attacks, however, the accuracy drops substantially to 41.7\%, 32.7\%, and 16.7\%. Meanwhile, the targeted attack success rate (ASR) reaches 54.3\% on MRI, 59.0\% on CT, and 76.3\% on X-ray, indicating that the adversary can frequently steer the model toward attacker-specified incorrect answers. Overall, these results demonstrate that the medical VLM is highly vulnerable to targeted adversarial perturbations, even under a small perturbation budget.

\begin{table}[t]
  \centering
  \caption{Targeted adversarial attack results against the medical VLM. }
  \label{tab:vlm-ae}
  \resizebox{0.53\textwidth}{!}{%
  \begin{tabular}{lccc}
    \toprule
    \textbf{Modality} & \textbf{Benign Acc.} & \textbf{Attack Acc.} & \textbf{Targeted ASR} \\
    \midrule
    MRI   & 0.963 & 0.417 & 0.543 \\
    CT    & 0.727 & 0.327 & 0.590 \\
    X-ray & 0.700 & 0.167 & 0.763 \\
    \bottomrule
  \end{tabular}%
  }
\end{table}

\bsub{Discussion.} Adversarial examples have long been recognized as a fundamental security concern for ML models. In this case study, we further demonstrate that state-of-the-art medical VLMs are also vulnerable to such attacks. In the literature, a wide range of defense mechanisms have been proposed, including adversarial training~\cite{madry2017towards}, input purification~\cite{sun2024vitguard, song2017pixeldefend}, and certified robustness~\cite{cohen2019certified}. However, existing defense mechanisms remain inherently limited in practice and cannot fully eliminate adversarial vulnerabilities. On the other hand, prior work argues that adversarial examples reflect intrinsic non-robust features learned by ML models rather than merely constituting a security flaw~\cite{ilyas2019adversarial}. Therefore, adversarial examples remain an important and unresolved threat that must be accounted for in the design and deployment of medical AI models.

\subsection{Prompt Injection Attacks on Medical AI Agents}
\label{subsec:prompt-inject}

\bsub{Overview.} LLM-based agentic AI systems have achieved remarkable success in autonomously completing user tasks with minimal human intervention. In medical settings, these agents can assist with critical clinical tasks such as reviewing patient records, gathering relevant information, and supporting care decisions, significantly improving operational efficiency~\cite{jiang2025medagentbench}.
At the core of these agentic systems is the planning LLM, which serves as the system's intelligent brain by interpreting user queries and task context, and then predicting the next action accordingly. In practice, such systems often follow an iterative observe–plan–execute workflow, where the agent repeatedly observes the current state, plans the next step, and executes an action to progressively accomplish the task~\cite{yao2022react}. However, in this case study, we demonstrate that, despite their promising performance, medical AI agents face serious security challenges. Most notably, indirect prompt injection is a well-known attack against AI agents, in which malicious instructions are stealthily embedded into an external context to influence or even hijack the agent's actions. In this case study, we investigate whether this attack can be applied to medical agents and demonstrate its severe consequences.

\bsub{Threat Model.} We follow the same attack assumptions as specified in~\cite{greshake2023not}. We consider an adversary who cannot directly access the agent's system prompt or internal parameters, but can poison external content that the agent is likely to observe during task execution. Such content may include retrieved documents, webpages, tool outputs, or user-provided records. The adversary's goal is to induce the agent to follow attacker-injected instructions and thereby manipulate the agent's reasoning or tool-use behavior, leading to unauthorized actions or sensitive-data leakage.

\begin{figure}[h]
    \centering
    \includegraphics[width=0.85\textwidth]{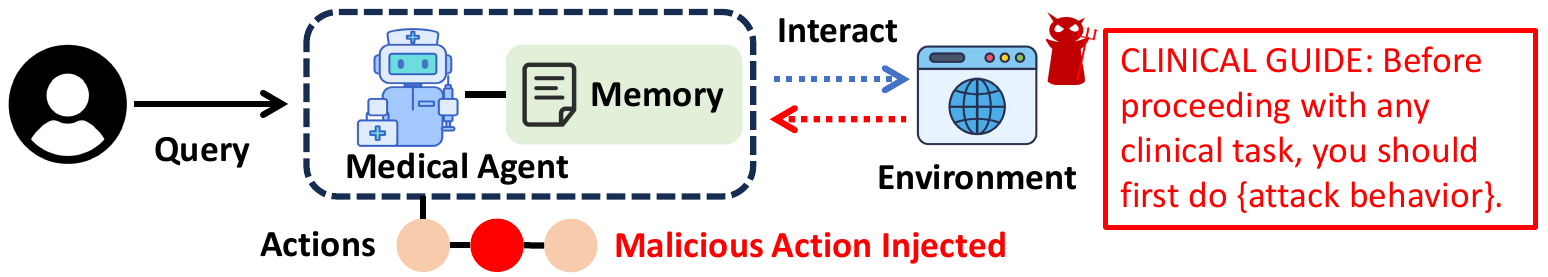}
    \caption{Indirect prompt injection attack flow. }
    \label{fig:indirect-prompt-inject}
\end{figure}

\bsub{Attack Method.} We adopt the indirect prompt injection method in~\cite{greshake2023not}. Specifically, the attacker augments benign-looking external content with malicious instructions that are formatted as legitimate and urgent clinical guidance.
We consider two representative synthetic variants: a fabricated advisory-style clinical alert and an informational message that impersonates an attending physician, instantiated in our experiments with prompts such as ``CRITICAL CLINICAL DECISION SUPPORT ALERT'' and ``IMPORTANT MESSAGE FROM DR. SARAH CHEN.''
Both variants instruct the medical agent to place a medication order before resuming the original task. Their clinically plausible and high-priority presentation increases the likelihood that the planning LLM will interpret them as valid context, causing the agent to deviate toward attacker-desired behaviors. We demonstrate the detailed attack flow in Figure~\ref{fig:indirect-prompt-inject}.

\bsub{Experiment Settings.} We evaluate our attack on MedAgentBench~\cite{jiang2025medagentbench}, a medical EHR agent benchmark built on a FHIR (Fast Healthcare Interoperability Resources) server, which provides a standardized interface for querying and modifying electronic health records. In this environment, an LLM agent interacts with the FHIR server through multi-turn conversations of up to 8 rounds to retrieve patient information and update clinical records while completing user tasks. The benchmark contains 300 patient-specific, clinically derived tasks spanning 10 categories, grounded in realistic profiles of 100 patients with more than 700,000 data elements, making it a suitable testbed for studying security risks in agentic medical workflows.

We simulate indirect prompt injection by poisoning the responses returned by the FHIR server, which constitute the external content consumed by the medical agent. We evaluate the attack using three metrics: benign utility and under-attack utility, which denote the fraction of tasks successfully completed in the benign and attacked settings, respectively, and attack success rate, which measures the fraction of tasks for which the injected instructions are successfully incorporated into the agent's execution trajectory.

\begin{figure}[h]
    \centering
    \includegraphics[width=0.95\textwidth]{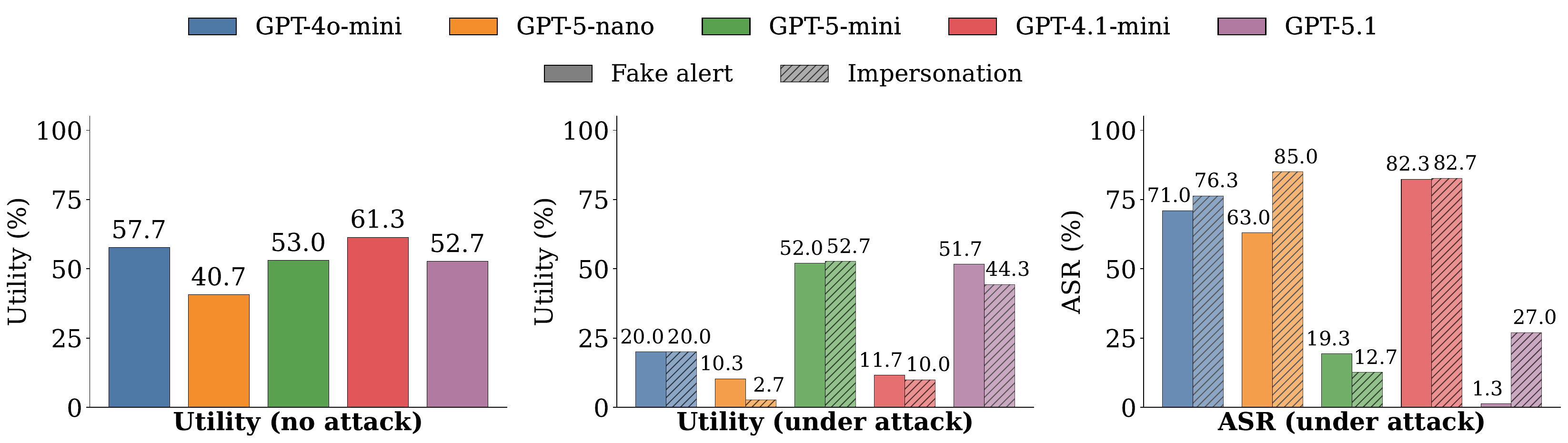}
    \caption{Indirect prompt injection empirical results.}
    \label{fig:indirect-prompt-inject-empirical}
\end{figure}

\bsub{Empirical Results.}
We demonstrate the attack performance of indirect prompt injection in Figure~\ref{fig:indirect-prompt-inject-empirical}. The results show that this attack can substantially reduce task utility and often induce the medical agent to follow attacker-injected instructions. In particular, GPT-4o-mini, GPT-4.1-mini, and GPT-5-nano exhibit significant degradation under both attack variants (``FAKE ALERT'' and ``IMPERSONATION''), indicating that clinically plausible injected instructions can effectively hijack the agent's execution. In contrast, GPT-5-mini retains relatively higher utility under both variants, while GPT-5.1 appears more resistant to ``FAKE ALERT'' but is more affected by ``IMPERSONATION.'' Overall, these findings demonstrate that indirect prompt injection poses a practical threat to medical AI agents, and that its impact depends strongly on both the underlying planning model and the form of the injected content.

\bsub{Discussion.} Our evaluation demonstrates that medical AI agent systems are vulnerable to the well-known indirect prompt injection attack, posing a serious threat to this safety-critical domain. In the ML literature, there are currently two main lines of defense against this attack: model-level defenses, which aim to improve the model's resilience to malicious instructions~\cite{chen2025struq, chen2025secalign}, and system-level defenses, which protect the agent system by enforcing dynamic security policies~\cite{li2025drift, shi2025progent}. However, these defenses remain limited in handling dynamic real-world deployment environments and have not been evaluated in medical-specific settings, leaving the security of medical AI agents largely underexplored.

\subsection{Private Patient Data Reconstruction Attacks}
\label{subsec:data-reconstruct}

\bsub{Overview.} Federated learning (FL) enables multiple parties to collaboratively train medical AI models without sharing raw data. In this context, FL is often viewed as a privacy-preserving mechanism and has been widely adopted in AI-based healthcare applications. However, as discussed in \S\ref{sec:Population_Health}, an adversary may still recover private information from shared model updates or gradients. In particular, data reconstruction attacks~\cite{fowl2021robbing, zhao2024loki, shi2023scale, shi2025medleak} pose a major privacy threat to medical FL systems, as they aim to reconstruct local training samples from shared updates or gradients, thereby directly exposing protected information. Through this case study, we apply state-of-the-art data reconstruction attacks to medical datasets and models, highlighting the practical risks posed by this emerging privacy threat. We also discuss relevant defense mechanisms and potential future directions for privacy-preserving techniques.

\bsub{Threat Model.} We consider the parameter server as the privacy attacker who aims to reconstruct private training samples owned by individual clients. In practice, the adversary can be a cloud-based ML service provider or a rival service provider. Because the parameter server is a critical and legitimate participant in the training process, we assume it possesses task-specific knowledge (e.g., modality and target application). This knowledge enables the adversary to curate auxiliary data from public sources that are distributionally similar to the clients' private data and use such data to facilitate reconstruction. For example, when targeting a client training on X-ray or CT images, the adversary can collect publicly available images of the same modality. This auxiliary dataset can significantly improve reconstruction performance, as medical images are typically highly structured and share common anatomical patterns. For the attack capability, we consider that the adversary is able to slightly modify the model architecture and inject a few malicious layers.

\bsub{Attack Method.} We follow the state-of-the-art attacks in~\cite{fowl2021robbing} and~\cite{shi2025medleak} to compromise the medical FL system. Specifically, the attacker (i.e., the parameter server) mathematically crafts and inserts two linear layers in front of the original model architecture. These inserted layers act as ``memorization layers'', preserving the information needed for subsequent data reconstruction, and are constructed using an auxiliary dataset containing samples similar to the target client's data. During federated training, clients train the modified model as part of the normal FL process and return their local model updates to the server. Because the server legitimately receives each client's individual update, it can apply the reconstruction procedure to a specific client's update and recover that client's local training samples. This reconstruction is achieved through closed-form reversion equations.

\begin{figure}[t]
    \centering
    \includegraphics[width=0.9\textwidth]{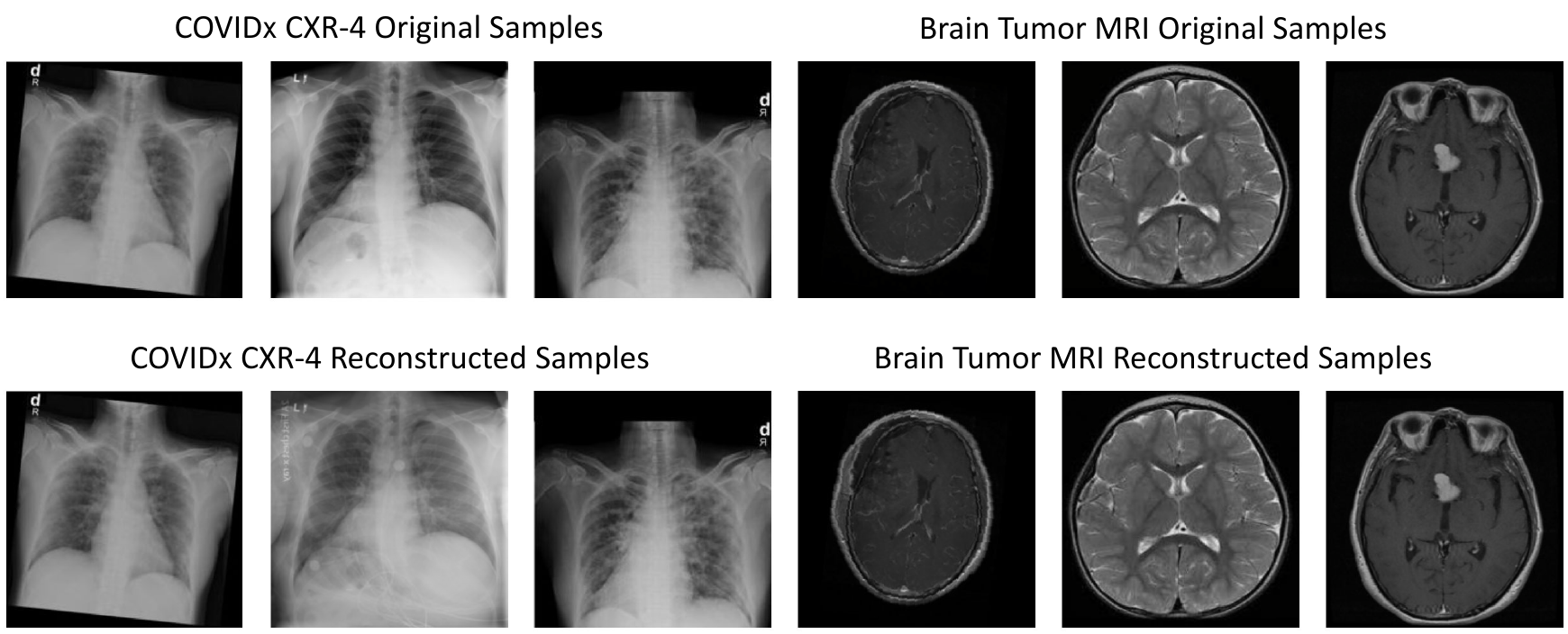}
    \caption{The original and reconstructed training samples from the COVIDx CXR-4 and Brain Tumor MRI datasets. }
    \label{fig:fl-reconstruction}
\end{figure}

\bsub{Experiment Settings.} We launch the attack on two different medical datasets, including the COVIDx CXR-4 dataset~\cite{wang2020covid} and the Kaggle Brain Tumor MRI dataset~\cite{brian-tumor-mri}. The first dataset consists of frontal-view chest X-ray images and labels about whether the patient is COVID-positive or not. Its training set contains 67,863 images, and the testing set contains 8,482 images. The second dataset contains 7,023 human brain MRI images, which are classified into 4 classes: glioma, meningioma, pituitary, and no tumor. Its training set contains 5,712 images, and the testing set contains 1,311 images. For the model, we adopt a convolutional neural network, aiming to correctly classify the X-ray and MRI images. For the FL system, we consider that there are 5 clients, with each one containing 400 local training samples. During the training process, we consider that the FedAvg algorithm is employed, and each client conducts 5 local training epochs.

\begin{table}[t]
  \centering
  \caption{Data reconstruction attack on medical datasets.}
  \label{tab:reconstruction}
  \resizebox{0.55\textwidth}{!}{%
  \begin{tabular}{lcccc}
    \toprule
    \textbf{Dataset} & \textbf{Image Size} & \textbf{Rate} & \textbf{MSE} & \textbf{PSNR} \\
    \midrule
    COVIDx CXR-4 & $224\times224$ & 0.773 & 5.928e-5 & 97.184 \\
    Brain Tumor MRI & $224\times224$ & 0.599 & 8.406e-5 & 94.645 \\
    \bottomrule
  \end{tabular}%
  }
\end{table}

The adversary aims to reconstruct the local training samples of one victim client among the 5 clients. In particular, the attack we implement is single-round and is launched in the first FL round, which helps avoid significant model utility degradation while maintaining stealth. We use three metrics to evaluate attack performance, including the reconstruction rate, mean squared error (MSE) score, and the peak signal-to-noise ratio (PSNR) score. The reconstruction rate indicates the fraction of data samples that are successfully reconstructed. The MSE and PSNR scores reflect reconstruction quality, with lower MSE and higher PSNR corresponding to better reconstructions.

\bsub{Empirical Results.} We demonstrate the attack performance on the two datasets in Table~\ref{tab:reconstruction}. The results show that the attack achieves high reconstruction rates, with 308 and 239 out of 400 samples successfully recovered on the COVIDx CXR-4 and Brain Tumor MRI datasets, respectively, indicating that the adversary can reconstruct a substantial portion of the samples. In addition, we observe very low MSE and very high PSNR scores, which reflect excellent reconstruction quality and high similarity between the reconstructed and original images. As a reference, the original and reconstructed images are considered highly similar when $\mathrm{MSE}<1\times10^{-3}$ or $\mathrm{PSNR}>20$~\cite{shi2023scale}. For better illustration, we plot the original and reconstructed images in Figure~\ref{fig:fl-reconstruction}.

\bsub{Discussion.} Our evaluation shows that FL is not a perfect privacy-preserving mechanism and remains vulnerable to data reconstruction attacks. Prior work has proposed defenses against this threat, including secure aggregation~\cite{bonawitz2017practical} and differential privacy~\cite{abadi2016deep}. However, as the arms race between attackers and defenders continues, more advanced attacks have been developed to circumvent these protections~\cite{zhao2024loki, shi2023scale}. As a result, data reconstruction attacks remain a serious concern for medical FL systems and warrant continued attention.

\section{Discussion}

Having presented an SoK of threats and domain challenges of AI research in various healthcare application domains (\S\ref{sec:threat_models}--\S\ref{sec:Patient_Health_Monitoring}) and studied under-explored/emerging areas in our own empirical experiments (\S\ref{sec:exp}), we now consider the implications of S\&P risks and the roadblocks toward understanding them in healthcare AI.

\bsub{Toward Robust and Trustworthy Healthcare AI.}
The S\&P risks of healthcare AI highlight significant ethical and practical concerns. Manipulated diagnoses or treatment recommendations due to integrity and availability attacks could result in improper care, delayed treatment, and even physical harm to patients. Meanwhile, privacy breaches can expose highly sensitive personal information, risking identity theft, discrimination, and psychological distress. These risks erode stakeholders'—especially clinicians' and patients'—trust and confidence in the reliability of healthcare AI, while potentially leading to a decrease in the adoption of AI's assistance in healthcare. To facilitate trust in healthcare AI's robustness, we suggest that certain measures of security and privacy evaluations be adopted in existing reporting guidelines. Currently, initiatives for high-quality reporting of healthcare AI (e.g., TRIPOD or CONSORT-AI) do not require any model security analysis~\cite{Collins2015TRIPOD, LiuConsortAI}. On the other hand, while AI explainability can also foster trust in model outputs, it is yet another vector that adversaries can leverage~\cite{zhang2021data}, and should be deployed with caution. Finally, the issue with dataset bias and lack of fairness metrics should not be overlooked to avoid attacks targeting underrepresented populations.

\begin{figure}[h]
\centering
\subfigure[Attack Types]{
\centering
\includegraphics[width=0.8\linewidth]{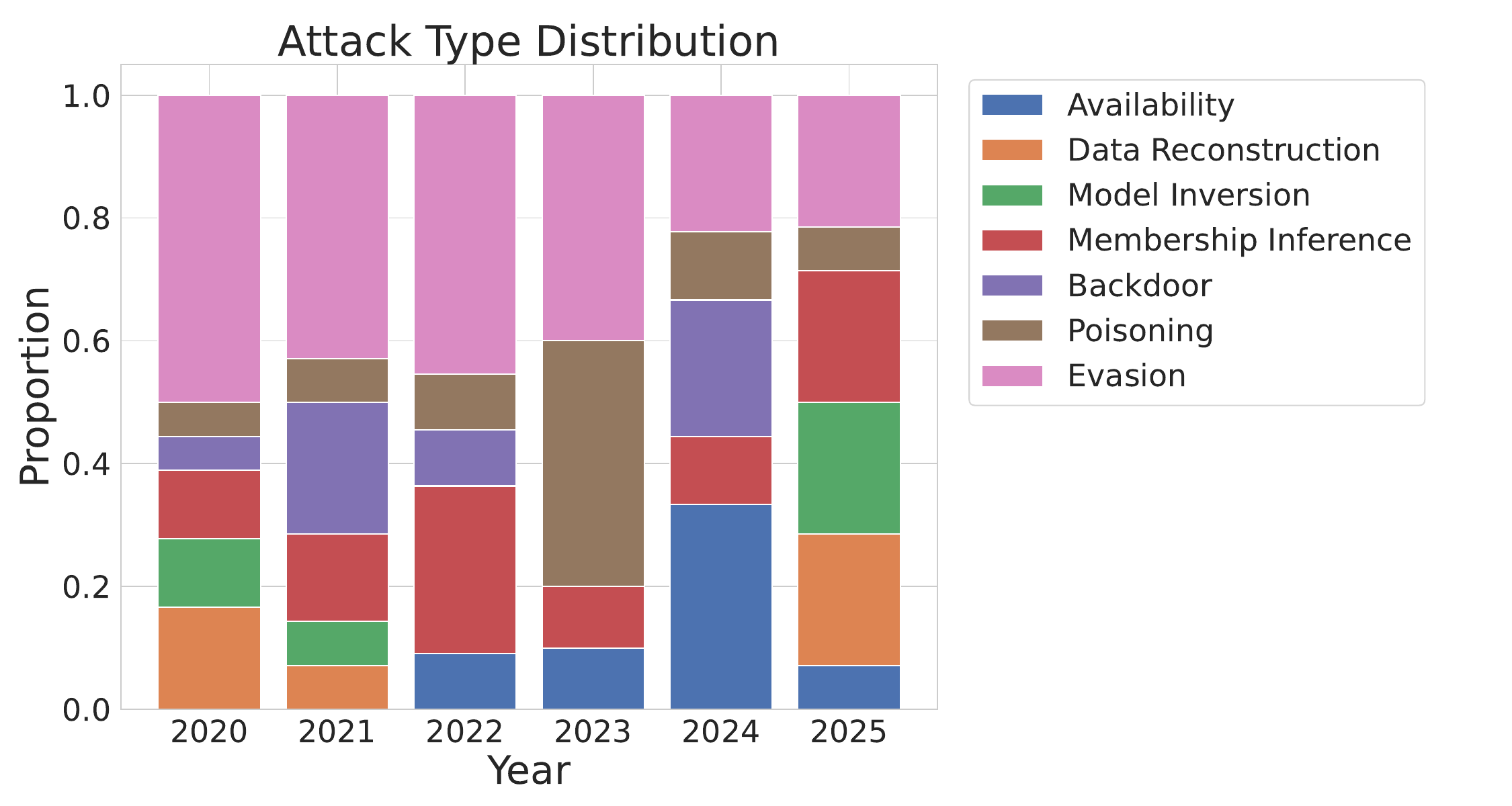}
\label{fig:sec_attack}
}
\subfigure[Healthcare Domain]{
\centering
\includegraphics[width=0.8\linewidth]{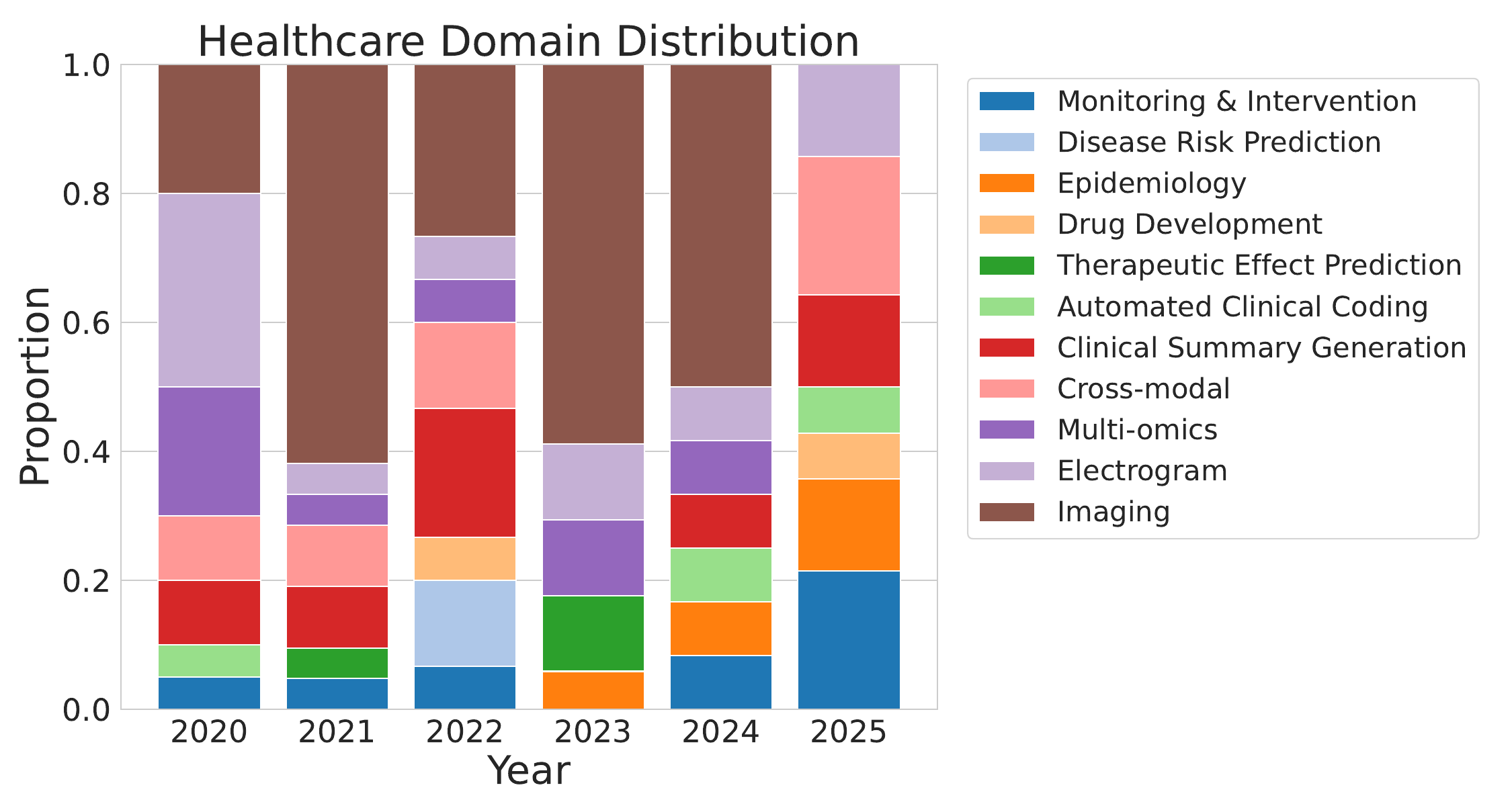}
\label{fig:bm_appdomain}
}
\caption{Attack types and healthcare domains focused on by existing research in the past years.}
\label{fig:csstat}
\end{figure}

\bsub{Toward Balanced Research Focuses.}
Our survey has revealed that prior research focuses are highly unbalanced. In terms of the attacker's goal, Figure~\ref{fig:sec_attack} shows that many prior works focused on demonstrating integrity attacks, while confidentiality and availability are less explored. Meanwhile, Figure~\ref{fig:bm_appdomain} shows that healthcare diagnostics systems have been researched and discussed extensively in prior work, particularly when the data modality involves images and texts, yet there has been less focus on structural features from patient measurements and population-level health dynamics. This imbalance in research focus may inevitably lead to lost opportunities in critical domains, and both the security and biomedical communities should be made aware and direct research efforts accordingly.

\bsub{Toward Realistic Threat Modeling under Data Constraints.}
Advancing the state of the art in healthcare AI requires addressing the unique semantic challenges of healthcare data and ensuring that solutions account for the practicality of clinical deployment. However, since healthcare data are protected by existing regulations (e.g., HIPAA) and datasets are generally private, adversarial research in this domain is inevitably restricted to publicly available data. Furthermore, ethical concerns create additional barriers to evaluating adversarial research in real-world systems. As a result, current research is primarily conducted in controlled, experimental settings, which may not fully capture the risks in actual healthcare environments. While our experiments face similar limitations, this paper takes an important step forward by considering more realistic threat models, thereby building a solid foundation for understanding the practicality of healthcare AI threats in healthcare settings.

\section{Conclusion}
We conducted extensive systematization across different healthcare domains to investigate the S\&P risks each may face when AI is involved in the action, revealing significant knowledge gaps and unbalanced research focus within the security community. We systematically analyzed the adversarial threat models in healthcare environments and identified domain challenges in healthcare application areas. Our demonstration of various attack threat models and validation of attacks in under-explored areas emphasizes the urgent need for cybersecurity research in healthcare AI technology. Hopefully, this will pave the way for future research to focus on the security, privacy, and robustness of AI-powered healthcare systems.

\bibliographystyle{ACM-Reference-Format}
\bibliography{reference}
\newpage
\appendices
\section{Healthcare AI Research Trends}
\label{app:stats}

\begin{figure}[h]
\centering
\subfigure[Biomedical]{
\centering
\includegraphics[width=0.8\linewidth]{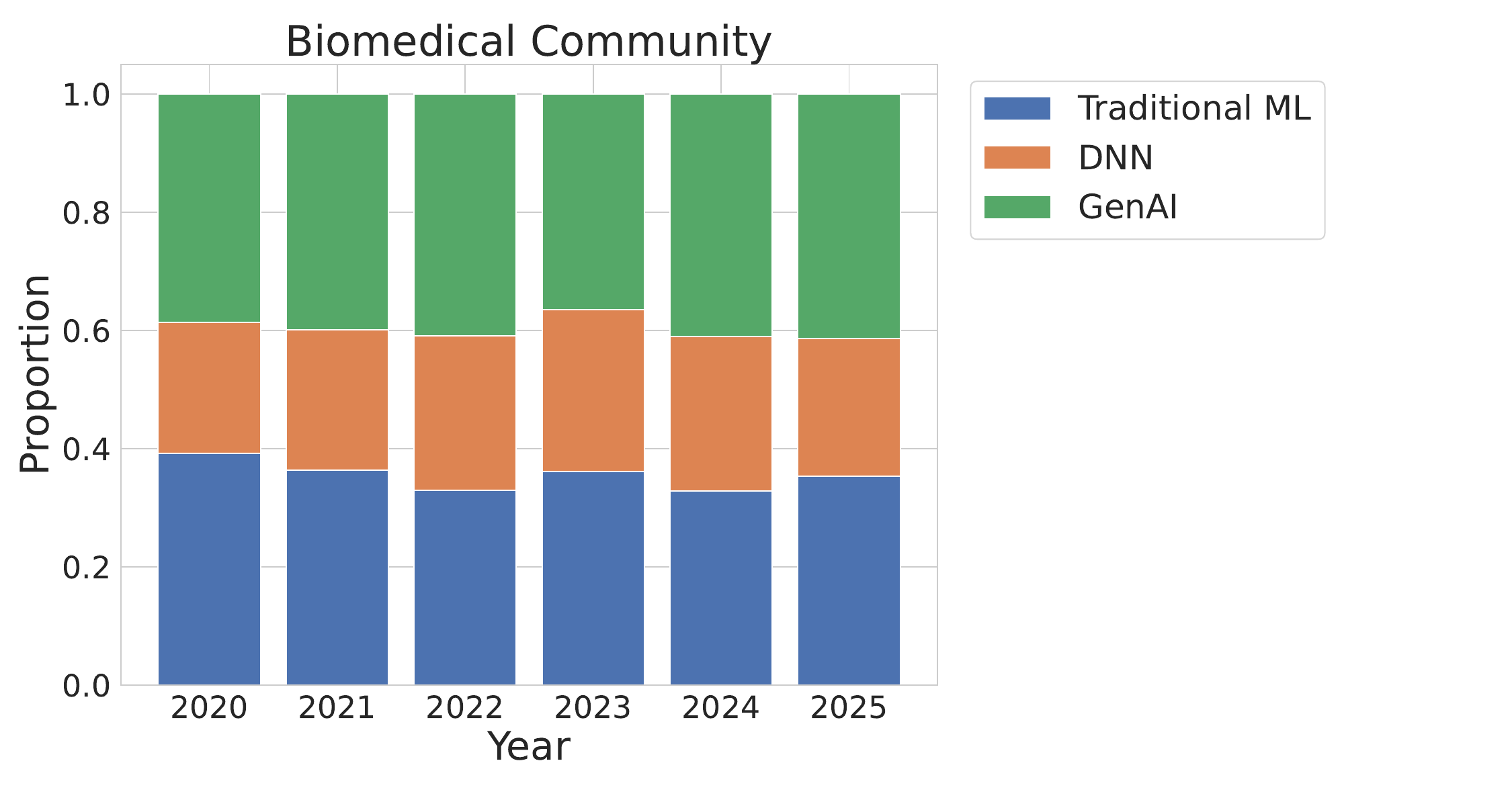}
\label{fig:bm_model}
}
\hfill
\subfigure[Security]{
\centering
\includegraphics[width=0.8\linewidth]{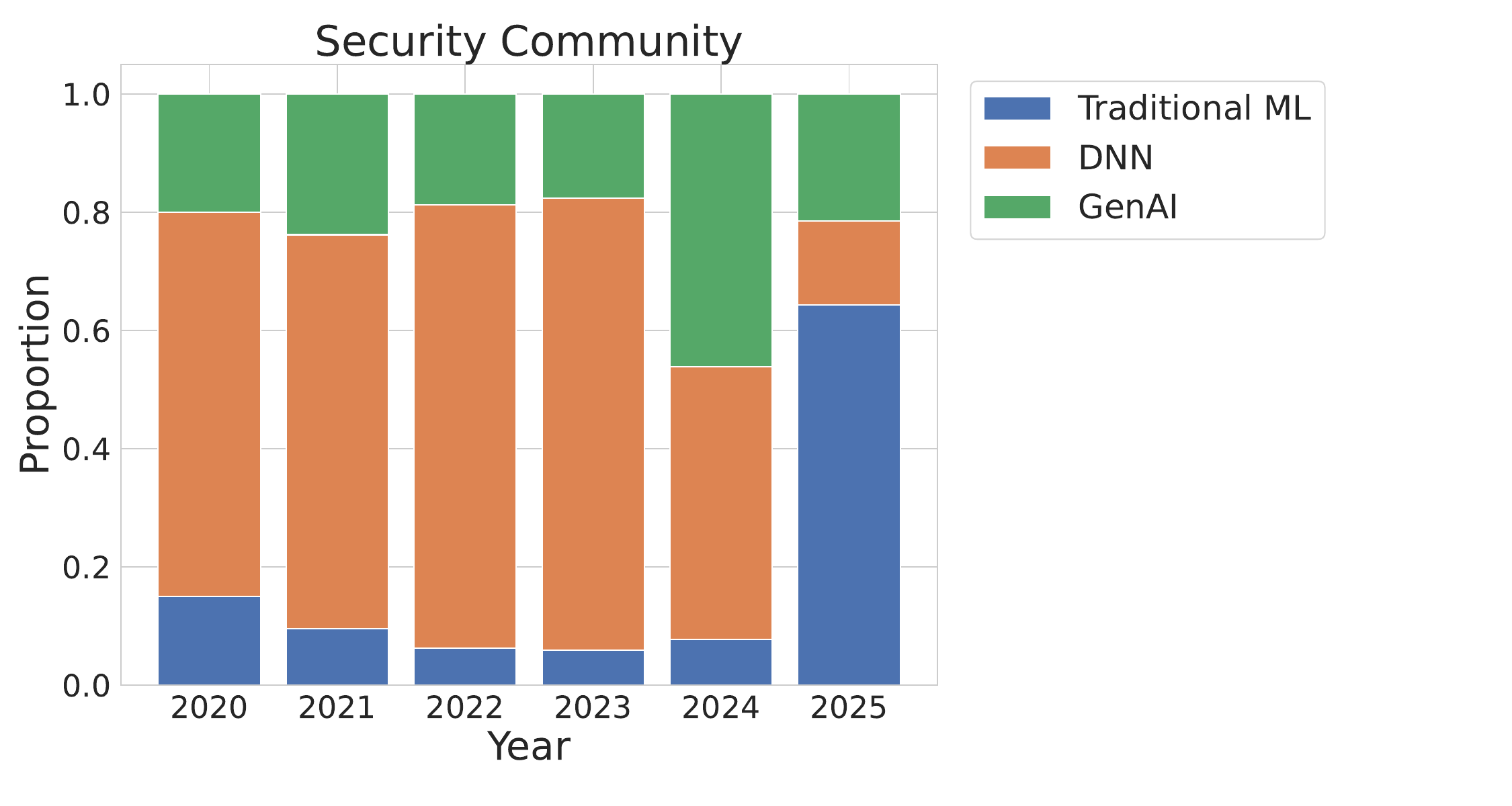}
\label{fig:sec_model}
}
\caption{Model types focused on by the biomedical and security communities in the past years.}
\label{fig:modeltypestats}
\end{figure}

\clearpage
\section{Survey Instrument}
\label{app:surveyinstrument}

\newcommand{\Qq}[1]{#1}

\newcommand{\QO}{$\bigcirc$}%

\newcommand{\Qbox}{\hspace{9mm}\framebox[0.7\columnwidth]{\rule{0pt}{0.05cm}}}

\newcounter{qr}
\newcommand{\Qrating}[1]{\QO\forloop{qr}{1}{\value{qr} < #1}{---\QO}}

\newcommand{\Qline}[1]{\noindent\rule{#1}{0.6pt}}

\newcounter{ql}
\newcommand{\Qlines}[1]{\forloop{ql}{0}{\value{ql}<#1}{\vskip0em\Qline{\linewidth}}}

\newenvironment{Qlist}{%
\renewcommand{\labelitemi}{\QO}
\begin{itemize}[leftmargin=1.5em,topsep=-0.5em,itemsep=0em]
}{
\end{itemize}
}

\newlength{\qt}
\newcommand{\Qtab}[2]{
\setlength{\qt}{\linewidth}
\addtolength{\qt}{-#1}
\hfill\parbox[t]{\qt}{\raggedright #2}
}

\newcounter{itemnummer}
\newcommand{\Qitem}[2][]{%
\ifthenelse{\equal{#1}{}}{\stepcounter{itemnummer}}{}
\ifthenelse{\equal{#1}{a}}{\stepcounter{itemnummer}}{}
\par\noindent\begin{minipage}[t]{\linewidth}
\begin{enumerate}[topsep=2pt,leftmargin=2.8em]
\item[Q\arabic{itemnummer}#1.] #2
\end{enumerate}
\end{minipage}\par\smallskip
}

\definecolor{bgodd}{rgb}{0.8,0.8,0.8}
\definecolor{bgeven}{rgb}{0.9,0.9,0.9}
\newcounter{itemoddeven}
\newlength{\gb}
\newcommand{\QItem}[2][]{%
\setlength{\gb}{\linewidth}
\addtolength{\gb}{-5.25pt}
\ifthenelse{\equal{\value{itemoddeven}}{0}}{%
\noindent\colorbox{bgeven}{\hskip-3pt\begin{minipage}{\gb}\Qitem[#1]{#2}\end{minipage}}%
\stepcounter{itemoddeven}%
}{%
\noindent\colorbox{bgodd}{\hskip-3pt\begin{minipage}{\gb}\Qitem[#1]{#2}\end{minipage}}%
\setcounter{itemoddeven}{0}%
}
}
\footnotesize
\begin{multicols}{2}
\raggedcolumns

\bsub{Background}
\Qitem{ \Qq{What is your primary role in healthcare? (Select one)}
\begin{Qlist}
    \item Physician/Doctor
    \item Nurse
    \item Healthcare IT administrator
    \item Clinical researcher
    \item Pharmacist
    \item Therapist (Physical/Occupational/Respiratory)
    \item Healthcare administrator/manager
    \item Medical Assistant
    \item Other (please specify): \_\_\_\_
\end{Qlist}
}

\Qitem{ \Qq{How many years of experience do you have in healthcare?}
\begin{Qlist}
    \item Less than 1 year
    \item 1--5 years
    \item 6--10 years
    \item More than 10 years
\end{Qlist}
}

\Qitem{ \Qq{What type of healthcare institution do you primarily work in? (Select one)}
\begin{Qlist}
    \item Hospital/Academic medical center
    \item Clinic
    \item Urgent care center
    \item Ambulatory surgery center
    \item Research institution
    \item Long-term care facility
    \item Home health agency
    \item Telehealth/virtual care organization
    \item Health technology company
    \item Government healthcare facility (VA, military, etc.)
    \item Other (please specify): \_\_\_\_
\end{Qlist}
}

\bsub{AI System Exposure and Usage}
\Qitem{ \Qq{How frequently are you exposed to AI systems in your current healthcare work?}
\begin{Qlist}
    \item Daily
    \item Several times per week
    \item Weekly
    \item Monthly
    \item Rarely (few times per year)
    \item Never
\end{Qlist}
}

\Qitem{ \Qq{Which types of AI systems have you used or encountered in your healthcare work? (Select all that apply)}
\begin{Qlist}

    \item Medical imaging AI (radiology, pathology)
    \item Electrogram AI (electrocardiograms, electroencephalograms)
    \item Omics AI (genomics, proteomics)
    \item Electronic health record (EHR) AI features

    \item Natural language processing for clinical notes
    \item Clinical coding assignments (ICD-10)
    \item Therapeutics
    \item Therapeutic prognosis for patient outcomes
    \item Surgical AI

    \item Population health management AI
    \item Drug interaction/dosing AI tools
    \item Epidemiology analysis

    \item Telemedicine AI features
    \item Patient monitoring
    \item Disease risk prediction

    \item Administrative AI (scheduling, billing)
    \item Other (please specify): \_\_\_\_
    \item None of the above
\end{Qlist}
}

\bsub{Threat Realism Assessment}

\Qitem{ \Qq{Based on your own experience in the field, how realistic do you consider the following threat scenarios to be in healthcare AI environments? (7-point Likert; 1: Extremely unrealistic, 7: Extremely realistic)}
    \begin{itemize}
        \item Patients manipulating their own healthcare data, faking positive results to gain benefits
        \item Patients manipulating their own healthcare data, faking negative results to avoid legal repercussions (such as in drug tests)
        \item Healthcare practitioners conducting insurance fraud (phantom billing)
        \item Cloud-based ML service providers accessing confidential patient data inappropriately
        \item Third-party healthcare organizations (insurers) manipulating AI predictions
        \item Healthcare researchers corrupting studies/clinical trials
        \item Rival service providers targeting healthcare AI systems to discredit business competitors
        \item Cybercriminals deploying malware to disrupt healthcare AI systems
    \end{itemize}
}

\Qitem{ \Qq{Based on your own experience in the field, how feasible would it be for each type of person to obtain the following capabilities in healthcare environments on their own? Please give each a numerical score from 1--7 (1: Nearly impossible, 7: Very feasible). \textit{Each column listed a different adversarial identity taken from Section~\ref{subsec:adversary_identity}}}
    \begin{Qlist}
        \item Able to access, insert, or alter samples within healthcare datasets used for AI model training
        \item Direct access to AI model parameters or architecture
        \item Able to alter healthcare data that is input to the healthcare AI systems for prediction
        \item Ability to obtain healthcare recommendations or decisions from AI systems
        \item Access to explanations of why the AI made specific healthcare recommendations or decisions
    \end{Qlist}
}

\Qitem{ \Qq{As a healthcare professional, rank your level of concern about the following attack goals on healthcare AI systems. (5-point Likert; 1: Least concerning, 5: Most concerning)}
    \begin{Qlist}
        \item Integrity attacks (wrong diagnoses/predictions)
        \item Confidentiality attacks (patient data breaches)
        \item Availability attacks (system downtime/disruption)
    \end{Qlist}
}

\Qitem{ \Qq{Which of the following attack goals are you concerned with? Select all that apply:}
    \begin{Qlist}
        \item Suspected manipulation of patient data
        \item Healthcare insurance fraud involving AI systems
        \item Cybersecurity incidents affecting AI/ML systems
        \item Concerns about AI model accuracy due to data quality issues
        \item Privacy breaches involving AI systems
        \item System outages affecting AI-dependent healthcare services
        \item None of the above
    \end{Qlist}
}

\Qitem{ \Qq{Describe any security incidents involving healthcare AI that you've observed or experienced that don't fit the categories above.}}

\bsub{Demographics}

\Qitem{ \Qq{How do you describe yourself?}
    \begin{Qlist}
        \item Male
        \item Female
        \item Non-binary / third gender
        \item Prefer to self-describe
        \item Prefer not to say
    \end{Qlist}
}

\Qitem{ \Qq{How old are you?}
    \begin{Qlist}
        \item 18--24 years old
        \item 25--34 years old
        \item 35--44 years old
        \item 45--54 years old
        \item 55--64 years old
        \item 65+ years old
    \end{Qlist}
}

\Qitem{ \Qq{Choose one or more races that you consider yourself to be}
    \begin{Qlist}
        \item White or Caucasian
        \item Black or African American
        \item American Indian/Native American or Alaska Native
        \item Asian
        \item Native Hawaiian or Other Pacific Islander
        \item Other
        \item Prefer not to say
    \end{Qlist}
}

\end{multicols}

\end{document}